 \documentclass[final,5p,times,authoryear]{elsarticle}

\usepackage{amssymb}
\usepackage{amsmath}

\usepackage{microtype}  
\usepackage[utf8]{inputenc}  
\usepackage[T1]{fontenc}     

\usepackage{ulem}
\usepackage{bm}
\usepackage{color}
\usepackage{amssymb}
\usepackage{amsmath}
\usepackage{graphicx}
\usepackage{amsfonts}
\usepackage{slashed}
\usepackage{pstricks}
\usepackage{float}
\usepackage{array}
\allowdisplaybreaks
\usepackage{dcolumn}
\usepackage{epsf}
\usepackage{tabularx}
\usepackage{soul}
\usepackage{wrapfig}
\usepackage[colorlinks=true, linkcolor=red, citecolor=blue, filecolor=cyan, urlcolor=magenta]{hyperref}
\usepackage{aas_macros} 
\usepackage{comment}

\definecolor{darkgreen}{rgb}{0.0, 0.5, 0.0}

\journal{Nuclear Physics B}

\begin{document}

\begin{frontmatter}



\title{Triggering Electron Capture Supernovae: Dark Matter Effects in Degenerate White-Dwarf-like Cores of Super-Asymptotic Giant Branch Stars}


\author[1]{Vishal Parmar}
\ead{vishal.parmar@pi.infn.it}

\author[2]{Domenico Scordino}
\ead{domenico.sc94@gmail.com}

\author[1,2]{Ignazio Bombaci}
\ead{ignazio.bombaci@unipi.it}

\affiliation[1]{organization={INFN, Sezione di Pisa},%
            addressline={Largo B. Pontecorvo 3},%
            postcode={I-56127},%
            city={Pisa},%
            country={Italy}}

\affiliation[2]{organization={Dipartimento di Fisica ``E. Fermi'', Università di Pisa},%
            addressline={Largo B. Pontecorvo 3},%
            postcode={I-56127},%
            city={Pisa},%
            country={Italy}}



\begin{abstract}
Electron-capture supernovae (ECSNe) have emerged as a compelling formation channel for low-mass neutron stars, bolstered by decades of theoretical work and increasingly supported by observational evidence, including the recent identification of SN~2018zd. Motivated by this, we investigate the influence of fermionic asymmetric dark matter (ADM) on the equilibrium structure of progenitor cores and the formation of their neutron star remnants. Using a general relativistic two-fluid formalism, we model the coupled evolution of ordinary matter (OM) and ADM, treated as separately conserved fluids interacting solely through gravity. Our analysis focuses on neon-rich white dwarfs (Ne WDs), which are typical progenitor cores for ECSNe. We assume conservation of both baryon number ($N_B$) and DM particle number ($N_D$) during collapse, allowing for a consistent mapping between progenitor and remnant configurations. We find that ADM significantly enhances the central density of the WD progenitor. This lowers the threshold gravitational mass $M^*$ required to initiate electron capture, enabling ECSNe from lower-mass progenitors. The resulting remnants are stable, DM-admixed neutron stars with gravitational masses potentially well below current observational bounds. Moreover, we find that the conversion energy during the WD-to-NS conversion is also significantly reduced for higher ADM particle masses and fractions, suggesting that unusually low-energy ECSNe may serve as potential indicators of ADM involvement in stellar collapse.
\end{abstract}


\end{frontmatter}


\section{Introduction} \label{sec:intro}


Stars with zero-age main-sequence mass in the range $(8$ -- $10)\, \mathrm{M_\odot}$ (in the case of solar metallicity) evolve to form strongly 
electron-degenerate oxygen-neon-magnesium (ONeMg) cores and become super-asymptotic giant branch (SAGB) stars. 
If the mass of such an ONeMg core grows to a threshold value $M^* \simeq 1.36\, \mathrm{M_\odot}$, electron capture (EC) on $^{20}\mathrm{Ne}$ and on $^{24}\mathrm{Mg}$ nuclei take place and ignite O-Ne deflagration around the center which generate a so called electron capture supernova (ECSN) \citep{Hiramatsu2021, Miyaji_1980, Nomoto1982, Nomoto1987, Jones_2013, Doherty_2017, Tominaga_2013, Jones_2016, Guo_2024}. Several multidimensional hydrodynamical simulations of these events identify two potential outcomes: 
(i) an electron capture-induced core-collapse supernova (ccECSN) that forms a neutron star \citep{Jones_2016}, or 
(ii) a thermonuclear explosion that either results in an ONeFe white dwarf \citep{Jones_2019} or leads to the complete disruption of the star \citep{Schwab_2020}. 
The ccECSNe are expected to form  low mass ($M \sim 1.25\, \mathrm{M_\odot}$), low spin and low kick velocity neutron stars, forming a low mass peak 
in the neutron star mass distribution  \citep{Schwab_2010,Valentim_2011,Kiziltan_2013}. 

Until recently, no supernova had been unequivocally identified as originating from electron capture, largely due to uncertainties in theoretical predictions. However, supernova 2018zd (SN2018zd) was recently identified by  \cite{Hiramatsu2021} as the first robust evidence of an  ECSN. 
This event exhibited key characteristics consistent with ECSN models, including a super-asymptotic giant branch progenitor, circumstellar material enriched with nuclear burning products, a low-energy explosion, and a distinct nucleosynthetic signature. 
SN 2018zd thus provides a crucial observational benchmark for studying how electron captures on neon and magnesium initiate core collapse, 
improving our understanding of supernova mechanisms and the final evolutionary stages of transitional-mass stars. 

In spite of its many successes, Einstein’s theory of general relativity does not fully explain the observed kinematics of galaxies and galaxy clusters unless an additional, non-luminous component dark matter (DM) is introduced \citep{corbelli2000extended, nesti2023quest}. DM remains one of the most profound mysteries in fundamental physics, astrophysics, and cosmology \citep{cirelli2024darkmatter, ARBEY2021103865, bramante2025darkmattercompactstars}. 
Despite its lack of direct electromagnetic interactions, its gravitational effects are well-established across multiple scales, from galaxy rotation curves \citep{sofue2013rotation, pato2015dynamical}, to strong and weak gravitational lensing \citep{corbelli2000extended}, to the cosmic microwave background (CMB) and large-scale structure formation in the universe \citep{Peebels1982}. 
The standard $\Lambda$-cold DM ($\Lambda$CDM) model suggests that DM constitutes approximately 26\% of the total energy density of the universe, far exceeding the 5\% contributed by ordinary baryonic matter \citep{nesti2023quest}.

The inability of Standard Model particles to account for  DM has driven extensive searches for new physics, including weakly interacting massive particles (WIMPs), axions, sterile neutrinos, and macroscopic candidates like primordial black holes \citep{aprile2017first}. 
Despite decades of effort, DM remains elusive due to its presumed feeble interactions with ordinary matter, making direct detection experiments challenging. Indirect searches for annihilation or decay signatures in cosmic rays, gamma rays, and neutrinos have also yielded no conclusive evidence, while collider searches have placed stringent constraints on DM production \citep{Bertone2018}.  

Astrophysical compact objects such as neutron stars (NSs) and white dwarfs (WDs) offer an alternative laboratory for probing DM properties, leveraging their extreme densities, strong gravitational fields, and potential for DM capture and annihilation \citep{GP1998, Garnavich_1998}. 
The accumulation of DM within these compact objects could impact their structural properties (mass, radius, mass-shed frequency, etc.), their internal composition, their thermal evolution, and even lead to novel collapse scenarios, providing indirect but critical constraints on DM properties \citep{Kouvaris2010, Bramante2017}.  Additionally, DM’s role in early universe phenomena—such as inflation, baryogenesis, and phase transitions—remains an open question in cosmology. 
If DM interacts with other beyond-Standard Model sectors, its effects could be imprinted on the cosmic microwave background, large-scale structure formation, or gravitational wave signals from first-order phase transitions \citep{Schwaller2015, Caprini2016}. 
As experimental sensitivity continues to improve, multi-messenger approaches combining astrophysical, cosmological, and particle physics observations 
are expected to provide crucial insights into the nature of DM. 

NSs have been widely studied as potential DM laboratories due to their extreme densities and strong gravitational fields, which make them highly efficient at capturing DM \citep{Mariani_2023}. Their typically low temperatures also enhance their sensitivity as thermal detectors \citep{raj_2018, universe11030074}.  
Early research explored DM capture and subsequent thermal relaxation in NSs, but more recent studies have highlighted how DM scattering with Standard Model particles can lead to measurable overheating \citep{Baryakhtar_2017}. This effect, where DM transfers kinetic energy to NS constituents during its 
semi-relativistic infall, could be observed using next-generation infrared telescopes like the Thirty Meter Telescope (TMT)  and Extremely Large Telescope (ELT), with potential sensitivity to NSs within 100 parsecs of Earth. Additional studies using the James Webb Space Telescope (JWST) and radio telescopes such as FAST, CHIME, and SKA could provide further insights by identifying old, isolated NSs with anomalous thermal emissions. 

Although NSs have been extensively used in DM studies, WDs also offer a compelling avenue for investigation. 
DM interactions in WDs could manifest through various mechanisms, including excess heating due to DM annihilation \citep{bertone2008compact, McCullough_2010, Horowitz_2020, Cermeno_2018,Rafiei_2022, shahrbaf2025observationalprobesneutronstar}, gravitational collapse into a black hole that gradually consumes the WD \citep{Kouvaris_2011, Janish_2019}, 
potential triggering of thermonuclear explosions, or structural modifications if DM contributes significantly to the WD’s mass \citep{Bramante_2015, Leung_2013}. These diverse effects make WDs valuable complementary probes in the search for DM, offering additional constraints alongside NS-based studies.


The purpose of this manuscript is twofold. First, we explore the equilibrium structure of white dwarfs admixed with fermionic asymmetric DM (ADM) using a two-fluid framework that incorporates electron capture in a self-consistent manner. While such treatments have been well-established in neutron star studies 
(e.g. \citep{ellis2018dark, qcl7-m5kf, Barbat:2024yvi,SCORDINO2025371} for review, please see \citep{bramante2025darkmattercompactstars}), 
their application to white dwarfs remains limited. Previous studies, such as those by \citep{Leung_2013}, considered dense ADM cores in white dwarfs but did not account for the onset of electron captures. As a result, their configurations reach unphysical central densities ($\rho_{cOM}  \sim 10^{14} \, \mathrm{g \, cm^{-3}}$) for the ordinary matter fluid, more appropriate for neutron stars than for white dwarfs. 
In contrast, our analysis incorporates the electron-capture process as a limiting factor, yielding physically consistent white dwarf configurations up to the electron capture mass threshold.  
Second, we investigate the dynamical transition of such ADM-admixed ONeMg white dwarfs into neutron stars via the electron-capture supernova (ECSN) channel, which typically originates from SAGB progenitors in the  transitional-mass range $8$--$10\,\mathrm{M_\odot}$. 

In this work, to study the  WD $\rightarrow$ NS evolution through ECSN, we focus on neon-rich white dwarfs (Ne WDs) as our initial stellar configurations, motivated by their well-established role as electron-capture supernova progenitors. ONeMg cores near the Chandrasekhar mass undergo collapse initiated by electron captures, primarily on $^{20}$Ne and $^{24}$Mg~\citep{Nomoto_1984, 2015MNRAS.446.2599D}, although for simplicity we adopt the threshold condition corresponding to $^{20}$Ne only in our analysis. This scenario is further supported by recent hydrodynamical and MESA-based studies of accreting ONe WDs and double-WD merger remnants~\citep{Jones_2019, 2024ApJ...975..186Z}. 
We consider a scenario in which both the DM particle number $N_D$ and baryon number $N_B$ are conserved during the collapse of the degenerate ONeMg white-dwarf-like core into the remnant neutron star. This assumption implies that no accretion or ejection of either ordinary matter and DM components occurs during the transition, establishing thus a direct mapping between the progenitor configuration and the remnant.
This is in analogy with the scenario proposed in \citep{Bombaci-Datta_2000} to describe the conversion of a neutron star to a strange star. 
A key feature of our scenario is that the threshold gravitational mass $M^*$ required to trigger electron captures in the white-dwarf-like core is 
significantly reduced in the presence of ADM.  
This mechanism naturally leads to the formation of ultra-low-mass neutron stars, potentially with gravitational masses below the canonical minimum mass 
$M \sim 1\, \mathrm{M_\odot}$ predicted by ordinary neutron star models \citep{Haensel_2007_book}. 
This raises the intriguing possibility that some observed low-mass neutron stars could be the remnants of ADM-admixed ECSNe. 
Theoretically, this could have far-reaching implications for the formation and evolution of compact objects, potentially altering the mass distribution 
of neutron stars and providing new channels for ADM detection through astrophysical observations.

The article is structured as follows. In Section~\ref{sec:formalism}, we present the equations of state (EOS) for white dwarfs and neutron stars, incorporating lattice corrections, electron-capture thresholds.  We  describe the structure equations for DM-admixed compact stars. In Section~\ref{sec:results}, we discuss our numerical results. Finally, in Section~\ref{sec:summary}, we summarize our findings and highlight the key implications of our work.

\section{Formalism} \label{sec:formalism}
\subsection{White Dwarf Matter}

As discussed in the introduction, the progenitors of ECSNe evolve to form strongly degenerate, white-dwarf-like cores. 
Stellar evolution calculations indicate that these ONeMg cores typically consist of approximately 60\% $^{16}$O, 30\% $^{20}$Ne, and 10\% $^{24}$Mg by mass \citep{Jones_2016, Gutiérrez_2005}.
  { However, in the analysis that follows, we assume the white-dwarf-like core, from which the ECSN originates, consists of a single ion species denoted by $^A_ZX$, where $A$ and $Z$ denote the mass number and atomic number, respectively. For the present work, we consider a pure $^{20}$Ne white dwarf to model the core of SAGB stars which produce ECSNe. This simplification does not affect the qualitative results reported here, such as the reduction of the electron-capture threshold mass in the presence of DM admixture, though the exact numerical values of the threshold density and stellar mass could vary in a more detailed multi-component treatment.}

We consider the case of unmagnetized and zero temperature ($T = 0$) stellar matter, corresponding to white dwarfs where the  temperature is lower then the crystallization temperature \citep{Potekhin_2009}. 
It is also assumed that the atoms are fully ionized \citep{Haensel_2007_book}.  
Under these conditions, we assume that the ions form a regular crystalline lattice.  
The total energy density of this system can be written as 
\begin{equation}
\label{eq:energy_density}
\varepsilon = n_X\, M_N(A,Z)\, c^2 + \varepsilon_e + \varepsilon_L,
\end{equation}
The first term on the right-hand side of Eq. \eqref{eq:energy_density} represents the energy density contribution due to ions (i.e. nuclei, 
since we consider fully ionized atoms) and $n_X$ is  the number density of ions which is related to the number density of electrons $n_e$ by the charge neutrality condition $n_e = Z\, n_X$. 
The second term represents the energy density of electrons, modeled as an ideal relativistic Fermi gas at zero temperature, and can be expressed as (see e.g. \citep{Shapiro-Teukolsky_1983}): 
\begin{equation}
\label{eq:e_energy}
\varepsilon_e = \frac{m_e c^2}{8\pi^2 \lambda_e^3} 
\left[ x_e (1 + 2x_e^2) \sqrt{1 + x_e^2} - \ln\Bigl(x_e + \sqrt{1 + x_e^2}\Bigr) \right],
\end{equation}
here $m_e$ is the electron rest-mass, $x_e = \lambda_e\, {k_F}_e  = \lambda_e\, \left(3\pi^2 n_e\right)^{1/3}$ is the dimensionless Fermi momentum of the electron gas, and $\lambda_e = \frac{\hbar}{m_e c}$ is the reduced Compton wavelength of the electron.   
Finally the third term $\mathcal{E}_L$ is the Coulomb lattice energy, considering point like ions, expressed as 
in \citep{Chamel_2015, Lunney_2003} 
\begin{equation}
    \label{eq:col_energy}
    \varepsilon_L = C\, e^2 n_e^{4/3} f(Z),
\end{equation}
where, $C$ is the so called lattice structure constant given by 
\begin{equation}
    \label{eq:latter_constant}
    C = -\frac{9}{10} \Big( \frac{4 \pi}{3}\Big)^{1/3}
\end{equation}
and $f(Z)$ is a dimensionless function of the atomic number. For crystal lattices composed of only one species of ion, 
namely $^{A}_{Z}X$, it follows that $f(Z) = Z^{2/3}$ \citep{Jog_1982}. 
If in Eq. \eqref{eq:energy_density} we add and subtract the electron rest-mass energy density $m_e c^2\,n_e$ and if we neglect the total binding energy of the $Z$ electrons in the neutral atom $^A_ZX$ we can write $M_a(A,Z) = M_N(A,Z) + Z m_e$, where $M_a(A,Z)$ is the mass of the neutral atom $^{A}_{Z}X$. 
Consequently the total energy density of the WD material can be written as 
\begin{equation}
\label{eq:energy_density_2}
\varepsilon = \frac{M_a(A,Z)\,c^2}{Z}\, n_e + \varepsilon_e - m_e c^2 n_e  + \varepsilon_L.
\end{equation} 
The latter expression gives the total energy density in terms of the atomic masses whose measured values are typically tabulated 
(e.g.\citep{Lunney_2003}). 

The pressure of the system can be readily obtained using the thermodynamical relation $P=n_e^2\frac{d \varepsilon/n_e}{d n_e}$ 
and the expression for $P_e$ and $P_L$ can be written as:
\begin{align}
P_e &= \frac{m_e c^2}{8\pi^2 \lambda_e^3} 
\left[ x_e \left(\frac{2}{3} x_e^2 - 1 \right) \sqrt{1 + x_e^2} + \ln(x_e + \sqrt{1 + x_e^2}) \right], \label{eq:Pe} \\
P_L & = \frac{\mathcal{E}_L}{3}. \label{eq:Pl}
\end{align}
One can see that the ion mass play no role in the pressure of the system. 
While electrons can be treated as an ideal Fermi gas, small deviations can arise due to electron exchange and polarization effects \citep{Haensel_2007_book, Potekhin_2009, Chamel_2015}, in addition to the Coulomb corrections already accounted for (see Eq.~\eqref{eq:col_energy}). However, these additional corrections have been shown to contribute only a few percent of the Coulomb lattice energy density \citep{Chamel_2015}. 
As a result, their impact on the EoS of stellar matter is minimal and they are neglected in this work.

\subsection{Electron Capture Instability}
Inside the WD, at sufficiently high mass-densities, electron capture by a nucleus $^{A}_{Z}X$ can become energetically favorable 
\citep{Langanke_2021}:
\begin{equation}
\label{eq:ec}
{}^{A}_{Z}X + e^- \rightarrow {}^{A}_{Z-1}Y + \nu_e.
\end{equation}

Within the star, the pressure $P(r)$ must vary continuously \citep{Eddington_1926}, so that the process \eqref{eq:ec} occurs at constant pressure ($P^*$) rather than at constant density. In addition the EC process does not change the total number of nucleons $A$, and since the temperature is assumed to be fixed ($T=0$), the appropriate thermodynamic potential for analyzing the stability of the  WD matter is thus the Gibbs free energy per nucleon, 
given by: 
\begin{equation}
g = \frac{\varepsilon + P}{n},
\end{equation}
The pressure ($P^*$) at which the onset of electron capture, defined in \eqref{eq:ec}, takes place can then be determined by the condition that 
the Gibbs free energy per baryon remains unchanged before and after the electron capture reaction  \citep{Chamel_2015} 
\begin{equation}
    \label{eq:gibbs}
    g(A,Z,P^*)=g(A,Z-\Delta Z,P^*),
\end{equation}
where $\Delta Z =1$. Eq. \eqref{eq:gibbs} can be numerically solved to find the pressure $P^*$, and next to find the corresponding threshold 
mass-density $\rho^*$. 

\subsection{The equation of state of dark matter}
%
In the present work we consider non-self-annihilating fermionic DM. This so called fermionic asymmetric DM (ADM) \citep{KLZ_2009, Zurek_2014} carry a 
conserved charge which is analogous to the baryon number in the case of ordinary matter.
We denote this quantity as the DM particle number $N_D$. In the case of a truly stable, non-self-annihilating, asymmetric fermionic DM particle, the total particle number $N_D$ remains constant in an isolated system \citep{KLZ_2009, Zurek_2014} or within a time-scale where accretion or ejection of ADM can be neglected.  {The DM component is modeled as a cold, degenerate Fermi gas, a widely adopted approximation in both compact-object and galactic contexts (e.g., \citealt{Persic_1996, Domcke_2015}). The condition for degeneracy is robustly satisfied under stellar conditions, since the Fermi temperature of the DM fluid exceeds the core temperatures of white dwarfs by several orders of magnitude. }

We describe the ADM fluid as a non-interacting (ideal) gas of fermions with mass $m_\chi$ and spin $1/2$. 
The corresponding EOS is thus given by Eq. \eqref{eq:e_energy} and \eqref{eq:Pe} replacing the electron rest mass $m_e$, the adimensional electron Fermi momentum $x_e$ and the electron reduced Compton wave length $\lambda_e$ with the corresponding quantities $m_\chi$, $x_\chi$ and $\lambda_\chi$ for the DM Fermi gas. 

\subsection{The equation of state for neutron star matter}
\label{sect:NS_EOS}

In this work we model the ordinary matter (OM) fluid of a DM admixed neutron star (DANS) as a uniform electric-charge-neutral fluid of neutrons, protons, electrons, and muons in equilibrium with respect to the weak interaction ($\beta$-stable nuclear matter) at zero temperature ($T = 0$). 

Recently a new microscopic EOS for this system has been obtained by \citet{BL2018} (hereafter the BL EOS) for the zero temperature case, 
using the Brueckner-Hartree-Fock (BHF) quantum many-body approach (see \citep{BL2018} and references therein) starting from modern two-body and three-body nuclear interactions derived within chiral effective field theory (ChEFT) (e.g. \citep{2011PhR...503....1M, 2020RvMP...92b5004H}). 
The BL EOS reproduces the empirical saturation point (i.e. saturation density $n_{0} = 0.16 \pm  0.01~{\rm fm}^{-3}$, and energy per nucleon 
$E/A|_{n_0} = -16.0 \pm 1.0~{\rm MeV}$) of symmetric nuclear matter, and other empirical properties (symmetry energy $E_{sym}$ and its slope parameter $L$, incompressibility) of nuclear matter at the saturation density $n_0$ (see Tab. 2 in \citep{BL2018}). 

When computing static ordinary neutron star configurations the BL EOS (for the $\beta$-stable case) gives \citep{BL2018} a maximum mass $\rm{M}_{\rm max} = 2.08\, \mathrm{M_\odot}$, with a corresponding central density $\rho_c = 2.74 \times 10^{15}\, \mathrm{g/cm^3}$ and radius $\rm{R}(M_{\rm max}) = 10.26\, \mathrm{km}$  and a quadrupolar tidal polarizability coefficient $\Lambda_{1.4} = 385$ (for the $1.4\, \mathrm{M_\odot}$ neutron star \citep{Logoteta:2019cyb}) compatible with the constraints derived from GW170817 \citep{TheLIGOScientific:2017qsa}.
Recently, the BL EOS has been extended~\citep{Logoteta:2020yxf} to finite temperature and to arbitrary proton fractions. This finite-temperature EOS model has been applied to numerical simulations of binary neutron star mergers \citep{Bernuzzi:2020txg, Endrizzi:2018uwl, A.Prakash_2021}.  
Finally, to model the (ordinary matter) neutron star crust (i.e. for nucleonic density $\le 0.08$ fm$^{-3}$) we have used the Baym--Pethick--Sutherland \citep{bps71} and the  Negele--Vautherin \citep{NV73} EOS.

\subsection{Structure equations for DM admixed compact stars} 

Since the non-gravitational interaction between DM and ordinary matter is negligible compared to interaction which determine ordinary matter EOSs (e.g. \citep{bertone2005particle,aprile2017first}), it is possible to split the total energy-momentum tensor as the sum of the energy-momentum tensor of each of the two fluids (OM and DM) and to have covariant conservation for both of them.  
Accordingly, the equation of state of OM is independent on the state variables of DM and vice versa. In addition, it is assumed that each of the two fluids be a perfect fluid. Based on these assumptions, and further assuming a spherically symmetric and stationary distribution of OM and DM, the stellar structure equations in general relativity for DANS  and DM admixed white dwarf (DAWD)  take the following form  (see e.g. \citep{kain2020radial}), 
which generalizes the TOV equations to the case of two fluids interacting exclusively through gravity:  
\begin{eqnarray}
\frac{dP_j}{dr}  &=&  -G \, \frac{m_{tot}(r)\, \varepsilon_j(r)}{c^2 r^2}\, \Bigg( 1 + \frac{P_j(r)}{\varepsilon_j(r)}\Bigg) \nonumber\\
           &\times& \Bigg( 1 + \frac{4\pi r^3 P_{tot}(r)}{c^2 m_{tot}(r)}\Bigg)\, \Bigg( 1 - \frac{2\, G\, m_{tot}(r)}{c^2 r}\Bigg)^{-1}
\label{2fTOV_a}
\end{eqnarray}
and 
\begin{equation}
     \frac{d m_j(r)}{d r} = \frac{4 \pi}{c^2} r^2 \varepsilon_j(r) \,, 
\label{2fTOV_b}
\end{equation}  
where $G$ is the gravitational constant, $P_j$ and $\varepsilon_j$  (with $j =\, $OM,\,DM) are the pressure the and energy density for the OM and DM fluid, 
$m_j(r)$ is the gravitational mass enclosed within a sphere of radial coordinate $r$ (surface area $4\pi r^2$) for each of the two fluids, $m_{tot}(r) = m_{OM}(r) + m_{DM}(r)$ is the total gravitational mass enclosed within a sphere of radial coordinate $r$ and $P_{tot}(r) = P_{OM}(r) + P_{DM}(r)$ the total pressure.  

To solve the stellar structure equations \eqref{2fTOV_a} and \eqref{2fTOV_b} we need to specify the equation of state for the two fluids (see previous subsections) and the appropriate boundary conditions at the center ($r=0$) and at the surface ($r=R_j$, $j=OM, DM$) of the matter distribution for each fluid\,\cite{kain2020radial}:

\begin{equation*}
    m_j(0) = 0\,  \qquad   \varepsilon_j(0) = \varepsilon_{c,j} \,             
\end{equation*}
We define the radius $R_j$ of the distribution of fluid $j$ by the following condition
\begin{equation*}
    P_j(R_j) = P_j^{surf}
\end{equation*}
where $P_j^{surf}$ is a fixed value for the surface pressure of fluid $j$. For DM we use $P_{DM}^{surf} = 0$ while for 
neutron star ordinary matter we chose $P_{OM}^{surf} = P_{OM}(\rho_{OM}^{surf})$, where 
$\rho_{OM}^{surf} = 7.86\ \mathrm{g/cm^3}$ is the mass density of solid $^{56}\mathrm{Fe}$.  

For $r> R_j$ we define $P_j(r) = 0$. The radius of the star is defined as 
\begin{equation}
    R = \max \{R_{OM}, R_{DM}\}
\end{equation}
Integrating Eq.(\ref{2fTOV_b}) we get the total gravitational mass $M_j$ for each of the two fluids ($j = OM, DM$)
\begin{equation}
      M_j \equiv m_j(R_j) =  \frac{4 \pi}{c^2} \int_{0}^{R_j} r^2 \varepsilon_j(r) \, dr 
\label{Mgrav_i}
\end{equation}  
and the total gravitational mass of the DM admixed compact star (DACS) is
\begin{equation}
     M_{tot} = M_{OM} + M_{DM} \,.
\label{Mgrav_tot}     
\end{equation} 

\section{\label{sec:results} Results}
We now present the results of our investigation into the influence of DM on electron-degenerate WD-like cores of SAGB stars and its potential role 
in triggering electron-capture supernovae which forms neutron star remnants \citep{Jones_2019}. 

When the DM admixed ONeMg white-dwarf-like core of a SAGB  star grows up to the EC threshold gravitational mass $M^*$ 
an ECSN will occur and, in the case of a ccECSN \citep{Jones_2019}, will form a DM admixed neutron star remnant. 
In our scenario we assume that during the collapse of the DAWD-like core into the remnant DANS both the total baryon number $N_B$ and the total DM particle number $N_D$ are conserved. In other words, we assume that no accretion or ejection of either ordinary matter and DM components occurs during the stellar transition \citep{Bombaci-Datta_2000}. 
Thus to quantify the DM content of both the DAWD-like core and the remnant DANS we define the DM number fraction as:
\begin{equation}
    y_\chi = \frac{N_D}{N_B + N_D} \, ,
\label{DMfrac_Y}    
\end{equation}
where $N_B$  and $N_D$ can be computed as 
\begin{equation}
   N_{B}=\int_0^{R_{OM}} n_B(r)\, dV\, ,
\label{num_b}    
\end{equation}
\begin{equation}
   N_{D} = \int_0^{R_{DM}} n_{D}(r)\, dV\, . 
\label{num_d}    
\end{equation} 
In the expressions above $n_B(r)$ and $n_{D}(r)$ are respectively the  baryon number density and the DM number density and 

\begin{equation*}
    dV = \Biggl(1-\frac{2\, G\,m_{tot}(r)}{c^2r}\Biggr)^{-1/2}4\pi r^2dr
\end{equation*}
is the proper volume, where $m_{tot}(r) = m_{OM}(r) + m_{DM}(r)$ is the total gravitational mass enclosed within a sphere of radial 
coordinate $r$.
Thus within our scenario $y_\chi$ remains constant during the conversion of the EC threshold mass DAWD-like core to the remnant DANS.

The baryonic ($M_B$) and dark mass ($M_D$) of the DM admixed compact star (DACS) (DAWD or DANS) then follows as 
\begin{align}
    M_B = & m_n N_B\\
    M_D = & m_\chi N_D
\end{align} 
$m_n$ being the neutron's rest mass.  
$M_B$ represents the total rest-mass of the $N_B$ neutrons, dispersed at infinity, that form the ordinary matter component of the DM admixed compact star \citep{Bombaci-Datta_2000}. A similar interpretation holds for the stellar dark mass $M_D$ of the DACS. 

Notice that the DM fraction of a DM admixed compact star, usually defined in the literature as the ratio of the gravitational mass $M_{DM}$ of the DM fluid to the total gravitational mass of the star, i.e.
\begin{equation}
    f_\chi = \frac{M_{{DM}}}{M_{OM} + M_{DM}}\, ,
\label{DMfrac}    
\end{equation}
within our scenario will not be constant during the stellar process which forms the remnant DANS of an ECSN.

As stated before, in order to solve TOV equations for DACS we need to specify the two fluids central densities. One could also fix the DM fraction $f_\chi$ or $y_\chi$ and study star sequences along $f_\chi$ or $y_\chi$ level curve by varying the ordinary matter central densities. 
The DM accreted onto a compact object depends on the interaction cross section of the DM with ordinary matter, the local DM density (i.e. the position of the object in the galaxy) and on the accretion time. Also, one should consider DM accreted onto the progenitor star of the compact object during its life \citep{scott2009dark, sulistiyowati2014effects, lopes2019asteroseismology}. For neutron stars a description of the accretion rate is given in \citep{kouvaris2008wimp, del2020change} without taking into account DM self-interaction. Accretion of DM onto white dwarf has been studied, for example, in \citep{bell2021improved, bell2024heavy}. In this work we will not attempt to give a systematic description of the accretion of DM onto compact objects, instead we are going to consider DM fraction suitable for neutron stars as an upper limit for the DM content of a white dwarf (see our discussion on DM fraction in \citep{SCORDINO2025371}). Values we choose for $y_\chi$ produce corresponding $f_\chi$ values in line with those used, for example, in \citep{leung2015dark}. As we will show, as the DM particle mass increase, low values for $y_\chi$ are needed to produce relevant effects on the stellar structure of a white dwarf or neutron star.

\begin{figure*}[ht!]
    \centering
    \includegraphics[width=0.65\linewidth]{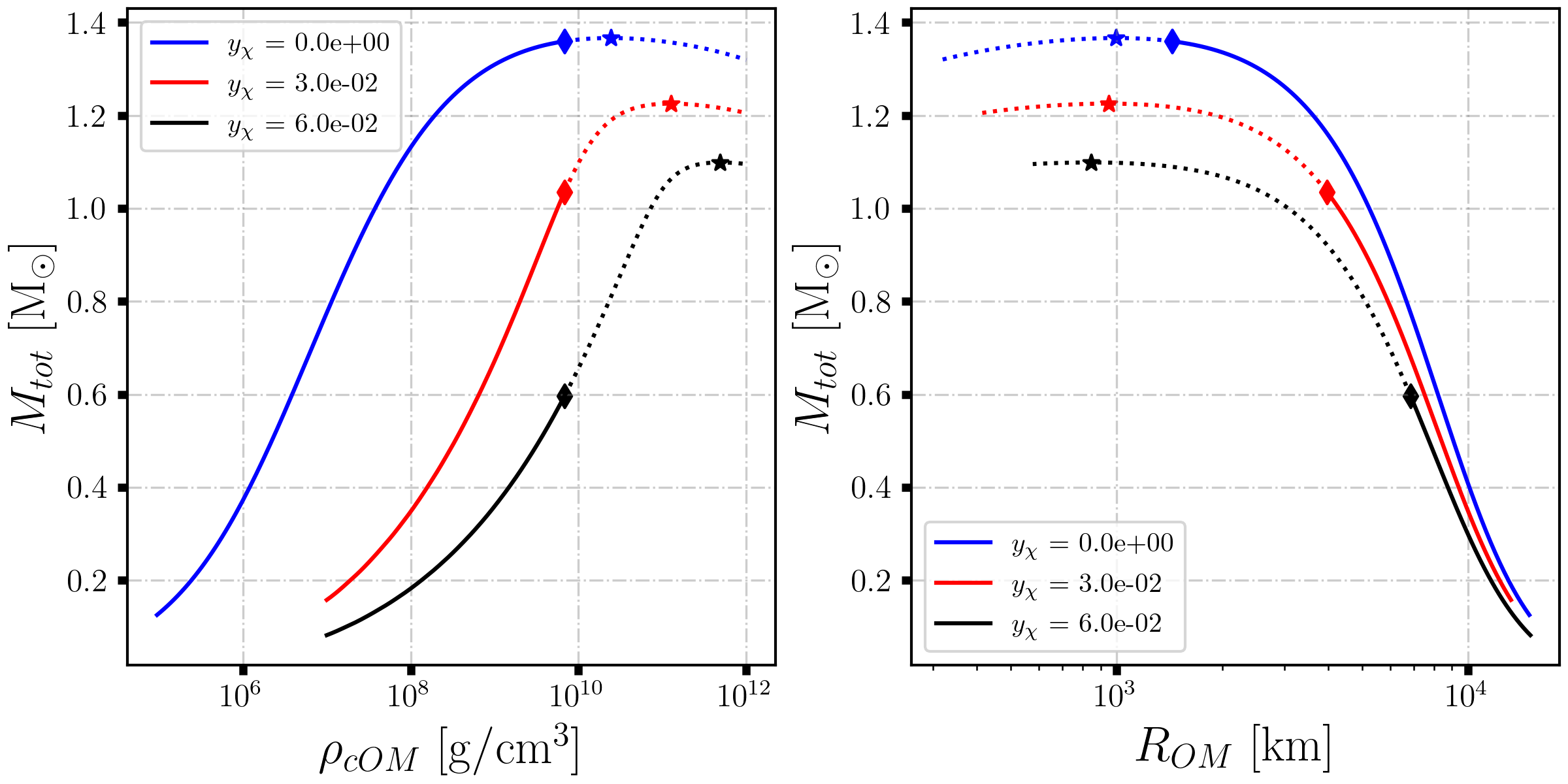}
     \includegraphics[width=0.65\linewidth]{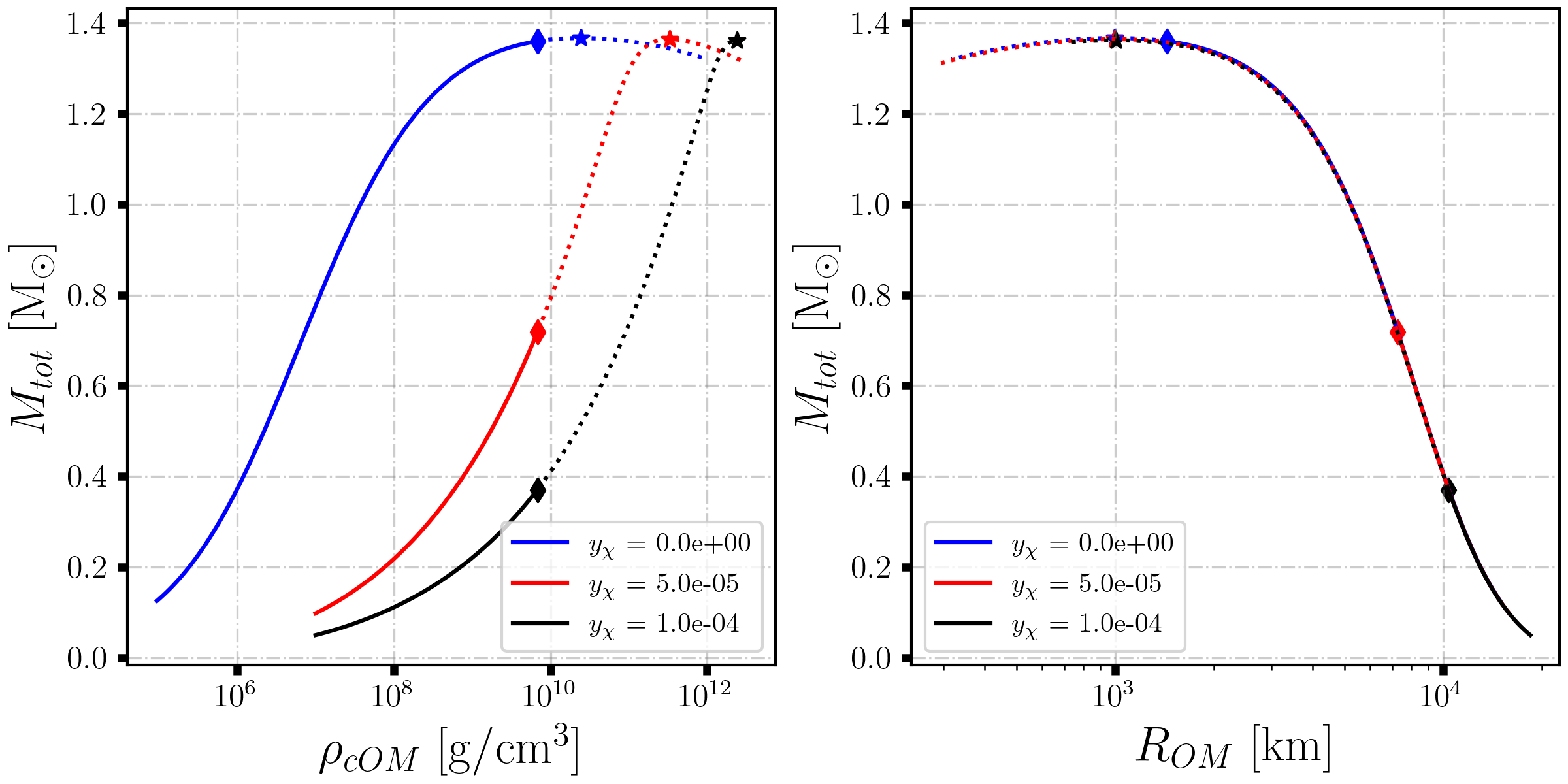}
    \caption{Total gravitational mass of the DAWD as a function of the ordinary matter central density $\rho_{c_{\rm OM}}$ (left panel) and as a function of 
     the radius $R_{\rm OM}$ of the ordinary matter distribution (right panel), shown for various values of the DM particle fraction $y_\chi$. 
     The results correspond to a pure $^{20}\mathrm{Ne}$ white dwarf. The top panels refer to $m_\chi = 1\,\mathrm{GeV}$, while the bottom panels are for $m_\chi = 10\,\mathrm{GeV}$. The diamond symbol on each line corresponds to the threshold mass $M^*(y_\chi)$ configuration for having electron capture in the center of the star, 
     while the star symbol denotes the the Chandrasekhar mass $M_{Ch}(y_\chi)$ configuration of the DAWD sequence.}
    \label{fig:M_R_rho}
\end{figure*}
\subsection{Effect of DM on degenerate white dwarfs}
\label{DM in WD}
%
 
In the present study, as stated earlier, we consider a pure $^{20}$Ne white dwarf to model the core of SAGB stars which produce ECSNe.  {In this work we adopt  WIMP masses of $m_\chi = 1$ GeV and $10$ GeV, values widely used in studies of DM admixed compact stars \citep{SCORDINO2025371, leung2015dark, Ivanytskyi_2020, kain2021dark, gleason2022dynamical, tolos2015dark, Das_2021, grippa2024constraintsscalarvectordark}. These choices highlight two qualitatively distinct regimes: light WIMPs in the few-GeV range that can form extended cores or halos in NSs, and heavier WIMPs producing small but dense cores. Current direct detection experiments place stringent constraints on the DM-OM interaction cross section ($\sigma_{\chi n}$) above $m_\chi \sim 10$ GeV, with XENON1T and LUX-ZEPLIN reporting limits of $\sigma_{\chi n} \lesssim 4 \times 10^{-47}$ -- $5 \times 10^{-48},\mathrm{cm}^2$ at $m_\chi \sim 30$ GeV \citep{aprile2017first, XENON,  LUX-ZEPLIN}. At lower masses, viable parameter space remains less constrained: \textsc{SuperCDMS} has excluded cross sections above $1.2 \times 10^{-42},\mathrm{cm}^2$ near $m_\chi \sim 8$ GeV \citep{SuperCDMS}, while \textsc{SENSEI} has probed down to the sub-GeV scale via electron recoils \citep{SENSEI}. Thus, our $m_\chi = 1$ GeV benchmark lies within the presently unconstrained light WIMP window, while $m_\chi = 10$ GeV sits at the edge of the strongly constrained region but remains illustrative for compact-star phenomenology. For completeness, we also briefly examine the qualitative impact of even lighter, sub-GeV candidates. Motivated by galactic-scale analyses of fermionic DM that successfully describe Milky Way and cluster dynamics in the keV–sub-MeV domain \citep{ARGUELLES201882, ARGUELLES2019100278}, we consider the representative case $m_\chi = 0.1\ \mathrm{GeV}$ to explore whether such particles could leave detectable imprints on the WD$\rightarrow$NS transition.}

In Fig.~\ref{fig:M_R_rho}, solving numerically the general relativistic two-fluid TOV equations \eqref{2fTOV_a} and \eqref{2fTOV_b}, we show the total gravitational mass $M_{tot}$ of the DM-admixed white dwarfs (DAWD) as a function of the central density $\rho_{cOM}$ of the ordinary matter component (left panels) and as a function the radius $R_{OM}$ of the ordinary matter distribution (right panels).  
The top panels correspond to a DM particle mass of $m_\chi = 1\,\mathrm{GeV}$, while the bottom panels show results for $m_\chi = 10\,\mathrm{GeV}$. 
Each curve represents a stellar sequence with a fixed value of the DM number fraction $y_\chi$.  
The solid portions of the curves indicate stellar configurations that are stable against electron capture, while the dashed portions denote configurations 
that are unstable.   
The diamond symbol on each curve marks the threshold gravitational mass $M^*(y_\chi)$ for electron capture, 
i.e. the DAWD configuration having a central density $\rho_{c\,OM}$ of ordinary matter equal to the threshold density $\rho^*$ for electron capture on the nucleus $^A_ZX$.  

In the case of electron capture on $^{20}\mathrm{Ne}$ 
\begin{equation}
     ^{20}\mathrm{Ne} + e^- \rightarrow{ ^{20}\mathrm{F} + \nu_e}
\end{equation}
$\rho^*(^{20}\mathrm{Ne}) = 6.82 \times 10^{9}\, \mathrm{g/cm^3}$ (see Table~\ref{tab:tab1}). 
Notice that the produced $^{20}\mathrm{F}$ undergoes subsequent a EC 
\begin{equation}
     ^{20}\mathrm{F} + e^- \rightarrow{^{20}\mathrm{O} + \nu_e}
\end{equation} 
having $\rho^*(^{20}\mathrm{F}) = 1.40 \times 10^{9}\, \mathrm{g/cm^3}\, .$ 
This further reduces electron pressure, thereby amplifying the WD instability that leads to an ECSN.

If we momentarily imagine turning off the weak interaction, specifically, by ignoring the possibility of EC on $^{20}\mathrm{Ne}$, we can integrate the two-fluid TOV equations up to the maximum mass configuration (the Chandrasekhar mass $M_{Ch}$) for each value of the DM number fraction $y_\chi$. 
The Chandrasekhar mass configuration is marked by a star symbol on each curve in Fig.~\ref{fig:M_R_rho}. 
Notice that in the present work WD (and DAWD) configurations are calculated using General Relativity and also including the Coulomb lattice correction to the 
EOS of the stellar material. Thus our calculated Chandrasekhar mass is lower than the "classical" Chandrasekhar mass derived in Newtonian gravity and for an ideal electron gas \citep{Carvalho_2018, Mathew_2017}.  

As seen from the results in Fig.~\ref{fig:M_R_rho}, the presence of DM significantly alters both the mass–radius ($M_{tot}$--$R_{OM}$)
and the mass–central-density ($M_{tot}$-$\rho_{cOM}$) curves. 
The properties of the Chandrasekhar mass configuration and of the EC threshold mass configuration relative to the DAWD stellar sequences shown in Fig.~\ref{fig:M_R_rho}, are summarized in Table~\ref{tab:tab1}. Our results indicate that the presence of DM in white dwarfs
leads to a reduction in the Chandrasekhar mass, to a decrease in the corresponding ordinary matter radius ($R_{OM,Ch}$), and an increase in the ordinary 
matter central density.  

\begin{table*}[tbh!]
\centering
\resizebox{.7\textwidth}{!}{%
\begin{tabular}{ccccccc}
    \hline
    \hline
    $y_{\chi}$ & $M_{ch}$ [$\mathrm{M_\odot}$] & $R_{ch} [\mathrm{km}]$ & $\rho_{c_{OM}}$ [$\mathrm{g/cm}^3$] & $\rho^*$ [$\mathrm{g/cm}^3$] & $M^*$ [$\mathrm{M_\odot}$] & $R^* [\mathrm{km}]$ \\ [1ex]
    
    \hline
    \multicolumn{7}{c}{No DM} \\
    \hline
    0  &  1.366  & 998.92 & $2.4201 \times 10^{10}$ &  $6.82 \times 10^9$ & 1.359 & 1440.7 \\[1ex]

    \multicolumn{7}{c}{$m_{\chi} = 1$ GeV} \\
    \hline
    $3  \times 10^{-2}$   &  1.225  & 956.05 & $1.2328 \times 10^{11}$ & $6.82 \times 10^9$ & 1.0353 & 3970.2 \\
    $6  \times 10^{-2}$   &  1.028  & 837.10 & $4.9770 \times 10^{11}$ & $6.82 \times 10^9$ & 0.5964 & 6865.6 \\
    \\ [1ex]
   
    \multicolumn{7}{c}{$m_{\chi} = 10$ GeV} \\
    \hline
    $ 5 \times 10^{-5}$  &  1.3634 & 1035.8  & $3.1257 \times 10^{11}$ & $6.82 \times 10^9$ & 0.7184 & 7272.6 \\
    $1 \times 10^{-4}$   &  1.3616 & 1005.13  & $2.3919 \times 10^{12}$ & $6.82 \times 10^9$ & 0.3707 & 10400.0 \\
    \hline
\end{tabular}} 
\caption{
Values of the DM number fraction $y_\chi$, Chandrasekhar mass $M_{Ch}$, the corresponding ordinary matter (OM) radius $R_{Ch}$ and central OM density 
$\rho_{c_{OM}}$, threshold density $\rho^*$ for the onset of electron capture, threshold mass $M^*$, corresponding OM radius $R^*$, for various values of the DM number fraction $y_\chi$.
}

\label{tab:tab1}
\end{table*}
\begin{figure*}[ht!]
    \centering
    \includegraphics[width=.45\linewidth]{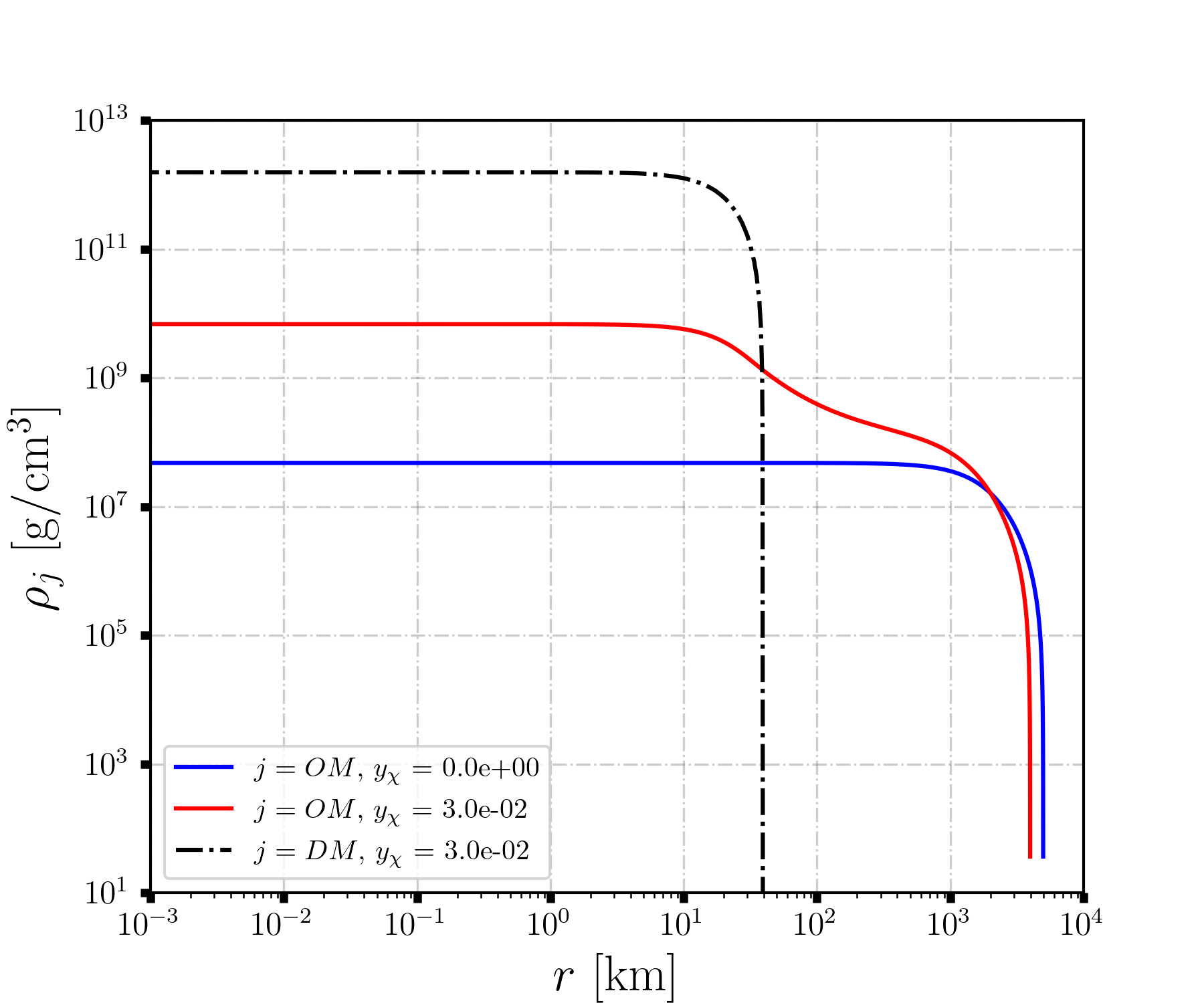}
    \includegraphics[width=.45\linewidth]{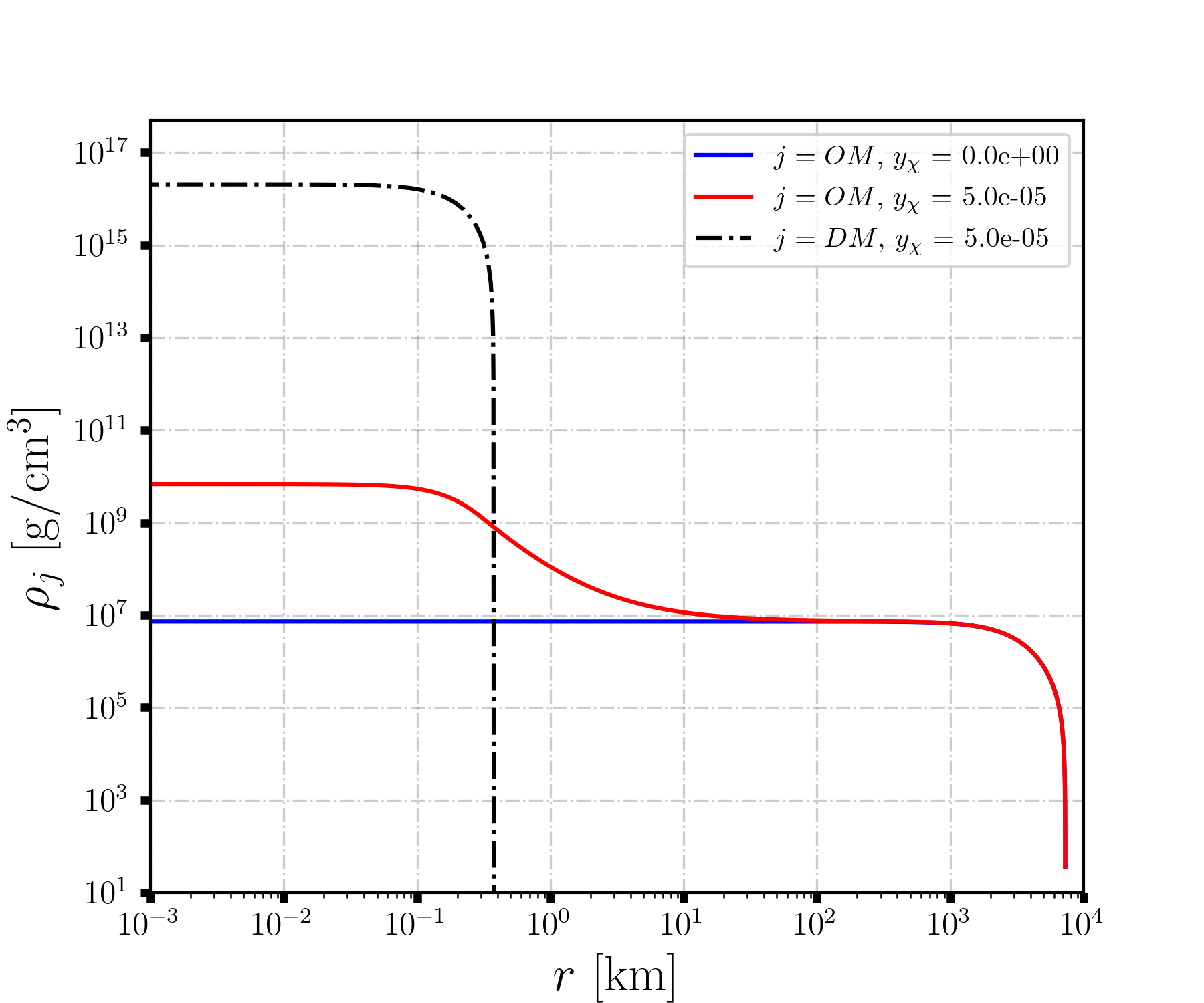}
    

\caption{Mass-density profiles $\rho_j(r)$ of ordinary white dwarf matter (solid lines) and DM (dash-dotted lines) for an ordinary ($y_\chi = 0$)   white dwarf (solid blue line) and a DM-admixed white dwarf. Both configurations have the same total gravitational mass, equal to the threshold mass for electron capture at a given DM number fraction $y_\chi$, i.e., $M_{\text{tot}}(y_\chi=0) = M^*(y_\chi = \text{const})$.
The left panel is relative to the case $m_\chi = 1\,\mathrm{GeV}$ and $y_\chi = 3\times 10^{-2}$ which results in $M^* = 1.035\, \mathrm{M_\odot}$. 
The right panel is relative to the case $m_\chi = 10\,\mathrm{GeV}$ and $y_\chi = 5\times 10^{-5}$ which results in $M^* = 0.718\, \mathrm{M_\odot}$. 
}
   \label{fig:density-profile}
\end{figure*}
\begin{figure*}[ht!]
    \centering
    \includegraphics[scale =0.45]{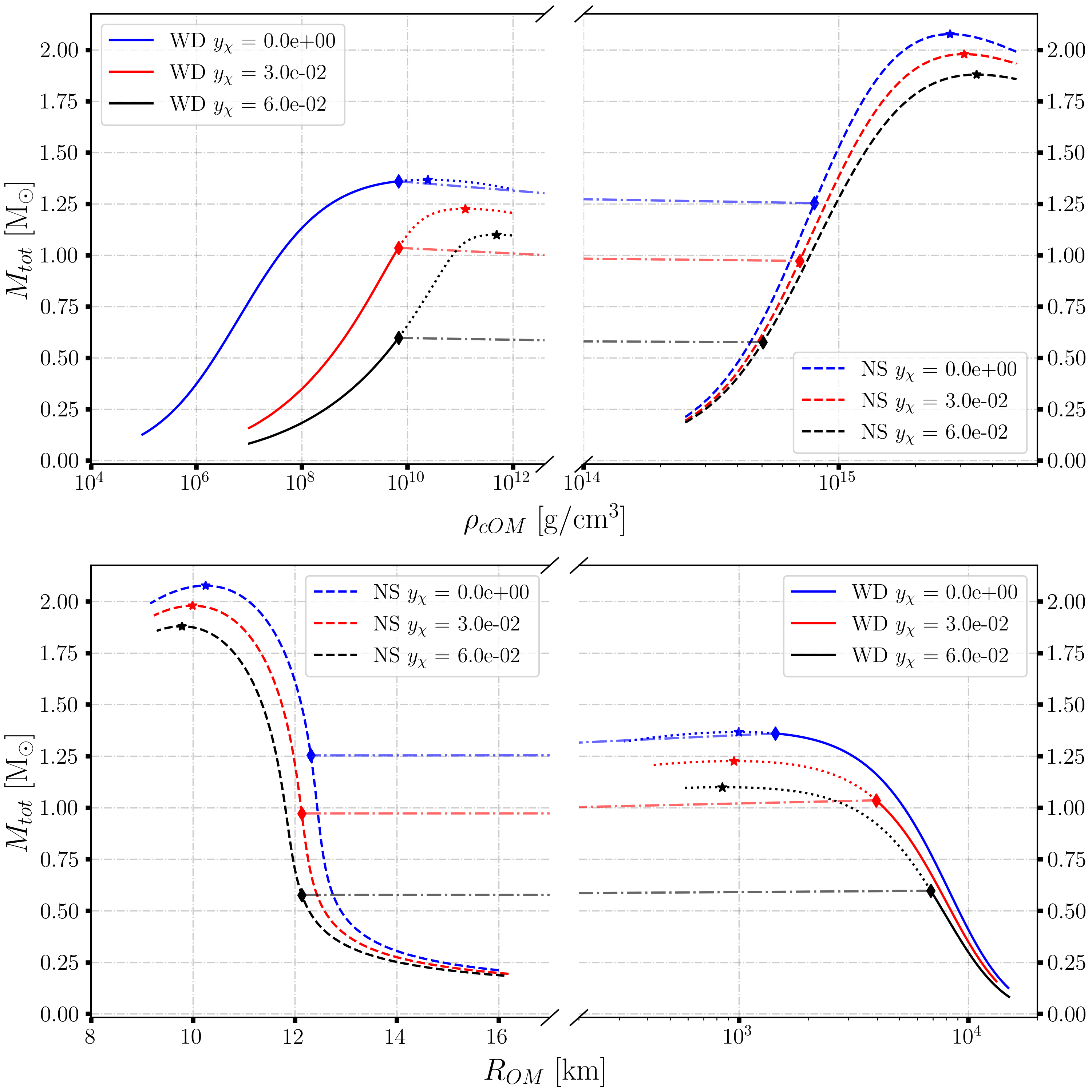}

        \caption{
        Stellar sequences for DAWDs (continuous lines) and DANSs (dashed lines) at fixed and equal DM number fraction $y_\chi$ (curves with the same color). The diamond symbol on each of the DAWD curves represents the threshold mass configuration $M^*(y_\chi)$ for electron capture. The corresponding diamond symbol (connected by the dash-dotted line) on the DANS sequence with the same $y_\chi$ represents the DANS remnant having the same baryon number $N_B$ (and thus the same DM particle number $N_D$) of the "ïnitial" EC threshold  mass DAWD configuration. Results are relative to a pure $^{20}\mathrm{Ne}$ ordinary WD matter and to a DM particle mass  
        $m_\chi = 1\, \mathrm{GeV}$.  
        } 
    \label{fig:wd_evl_m_rho_R_1}
\end{figure*}
\begin{figure*}[ht!]
    \centering
    \includegraphics[scale=.45]{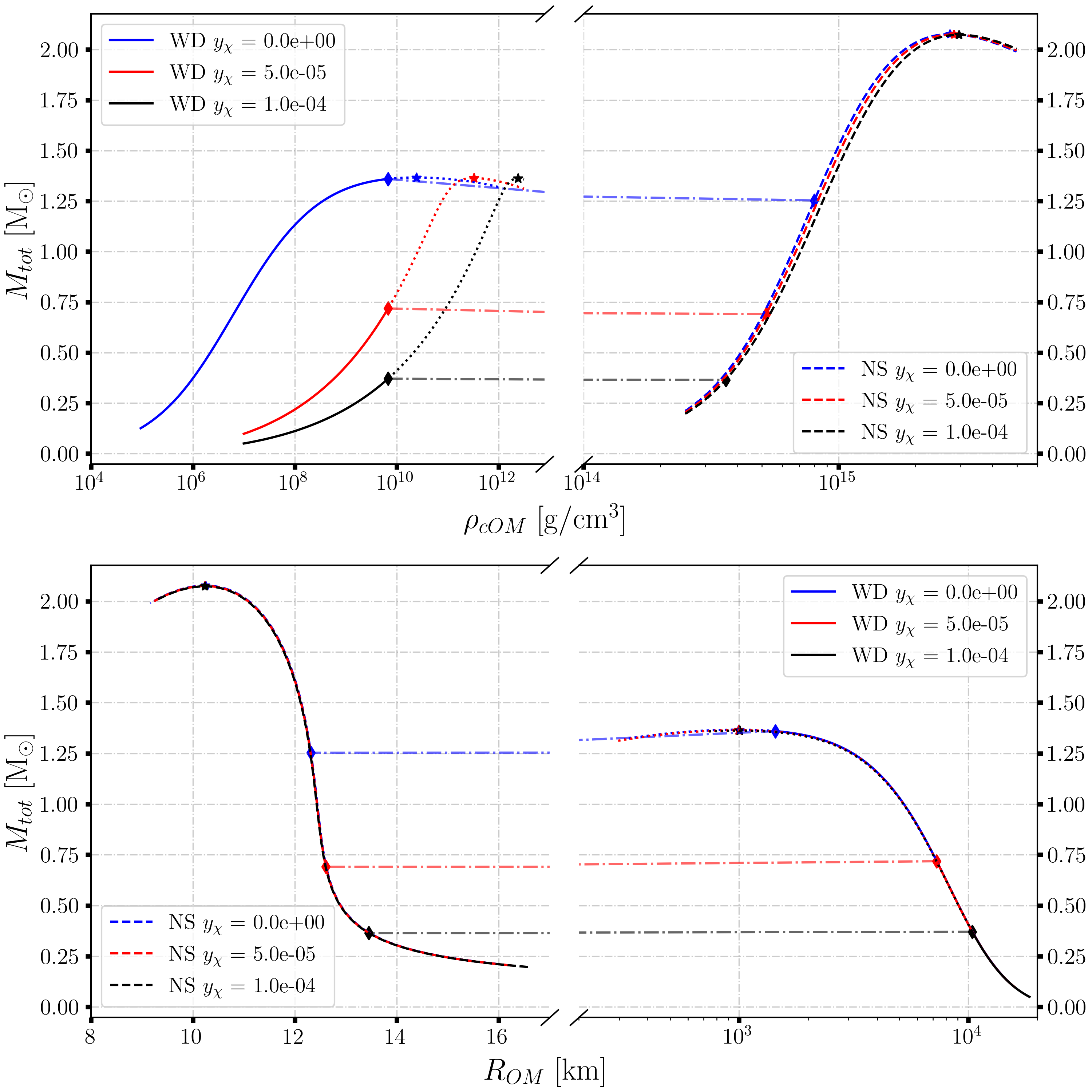} 
    
    \caption{Same as in Fig. \ref{fig:wd_evl_m_rho_R_1} but for \( m_\chi = 10\, \mathrm{GeV} \).}
    \label{fig:wd_evl_m_rho_R_10}
\end{figure*}

\begin{figure*}[ht!]
    \centering
    \includegraphics[width=.7\linewidth]{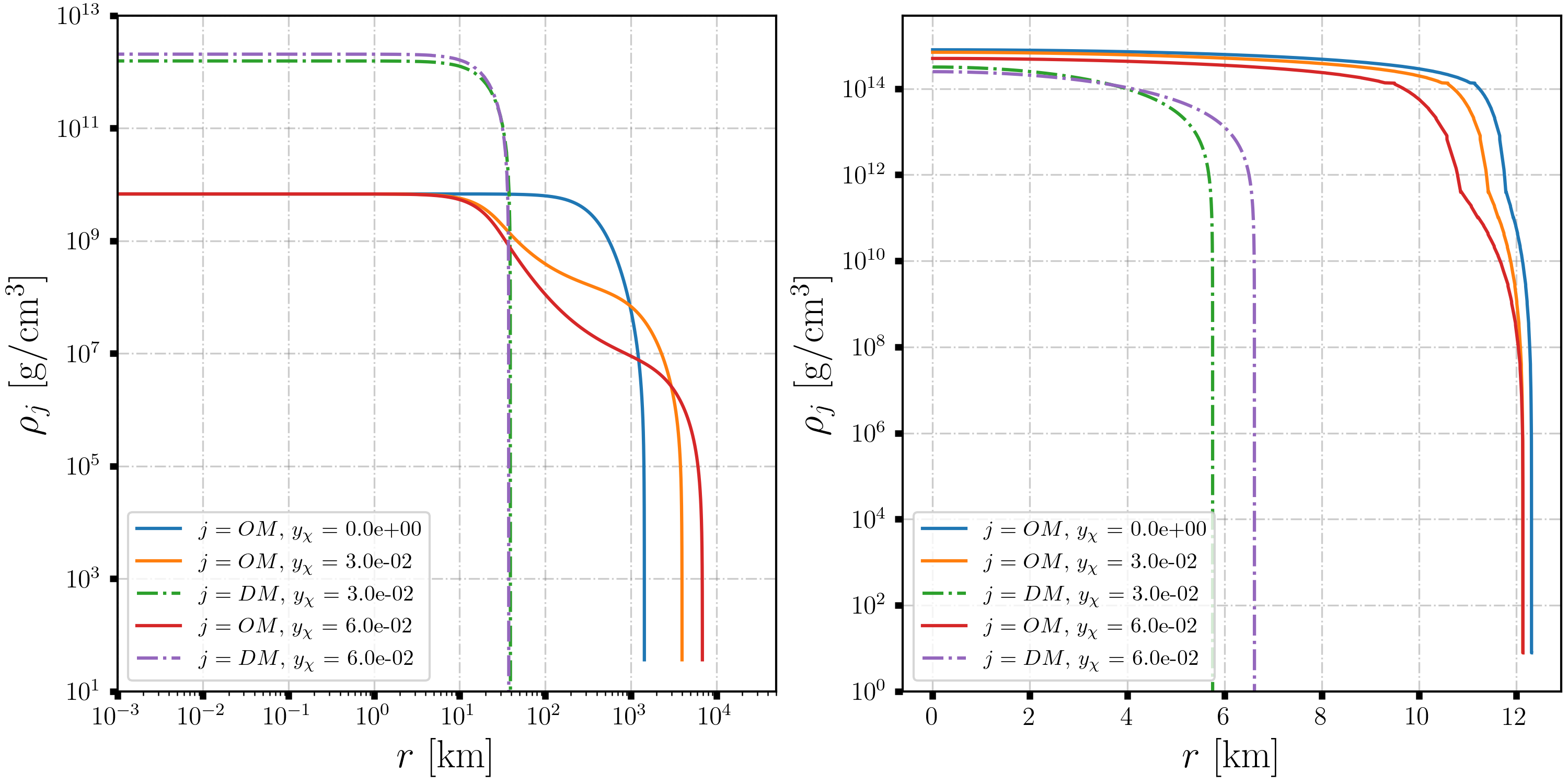}
    \includegraphics[width=.7\linewidth]{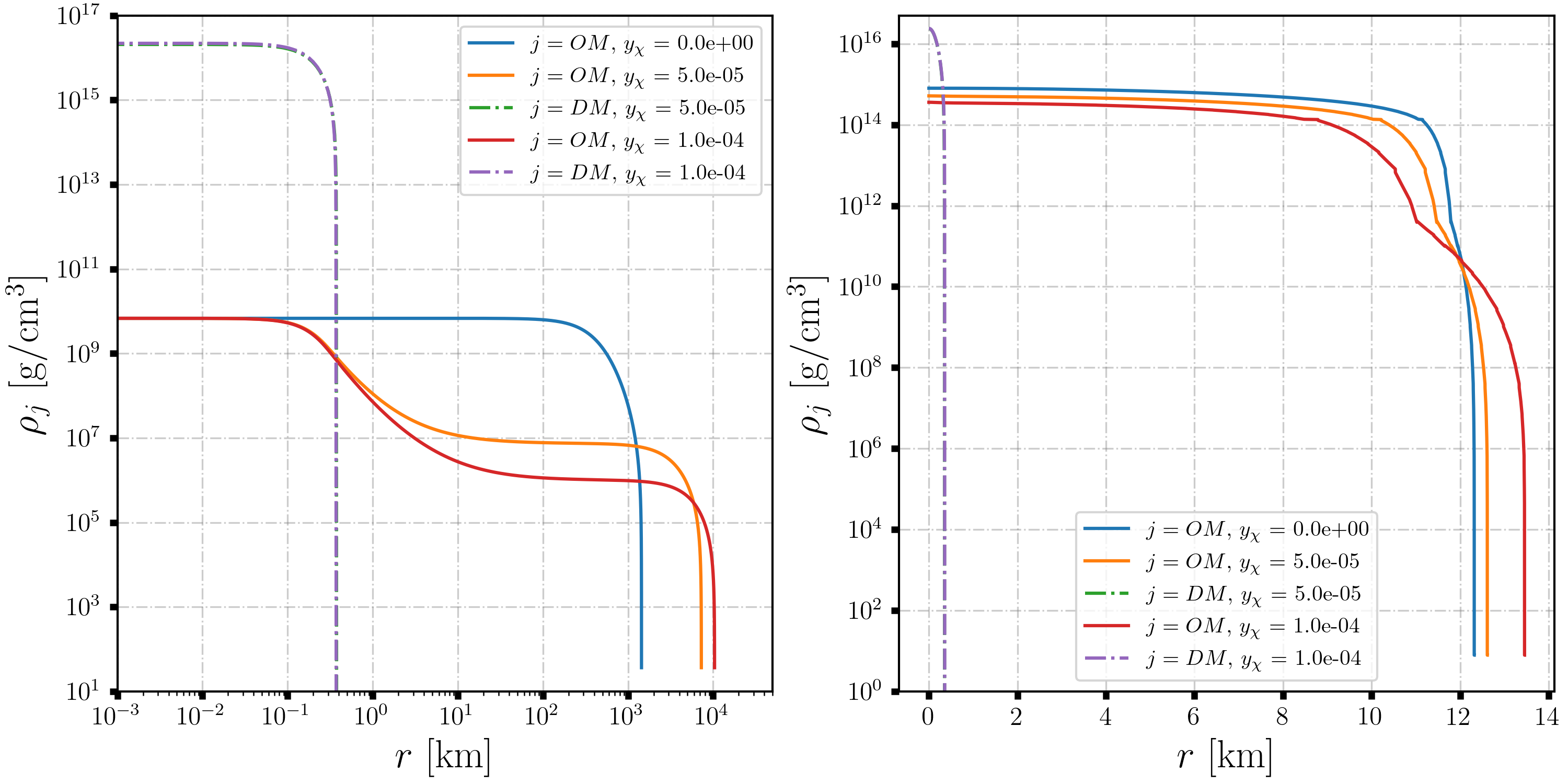}

    \caption{
Mass-density profiles $\rho_j(r)$ for ordinary matter (solid curves) and for DM (dash-dot curves) for the EC threshold mass DAWD configurations (left panels) and for the corresponding (curves with the same color) DANS remnants (right panels). 
The upper panels correspond to the case $m_\chi = 1\,\mathrm{GeV}$, while the lower panels correspond to $m_\chi = 10\,\mathrm{GeV}$. 
The values of the radii $R^*_{OM}$, $R^*_{DM}$, $R^{NS}_{OM}$ and  $R^{NS}_{DM}$ for the various mass-density distributions can be read in Table~\ref{tab:wd_ns_transition}. }

    \label{fig:dens_profile_WD-NS}
\end{figure*}

For the case with $m_{\chi} = 1~\mathrm{GeV}$, the Chandrasekhar mass $M_{Ch}(y_\chi)$ decreases from $1.366~\mathrm{M_\odot}$ (in the DM-free case) 
to $1.028~\mathrm{M_\odot}$ for a DM number fraction $y_{\chi} = 6 \times 10^{-2}$ (see Table~\ref{tab:tab1}). 
However, once the electron-capture threshold density $\rho^*$ is crossed at the center of the star, the white dwarf becomes unstable and 
undergoes gravitational collapse initiating the ECSN.   
Therefore, the stellar gravitational mass at the electron-capture threshold, denoted as $M^*$, becomes the relevant physical quantity for stability considerations. In other words $M^*(y_\chi)$ can be considered as the effective maximum mass configuration for the DAWD sequence. 

In the case with $m_{\chi} = 1~\mathrm{GeV}$, $M^*$ decreases from $1.359~\mathrm{M_\odot}$ (in the DM-free case) to $0.5964~\mathrm{M_\odot}$ in the presence of 
a 6\% DM number fraction ($y_{\chi} = 6 \times 10^{-2}$).   
As previously noted, an increase in $y_{\chi}$ leads to a slight decrease in the OM radius corresponding to the Chandrasekhar mass configuration $M_{ch}$.
In contrast, the OM radius $R^*_{OM}$ at the electron-capture threshold mass $M^*$ exhibits the opposite trend, showing a substantial increase due to the presence of  DM. For $y_{\chi} = 6 \times 10^{-2}$, this radius expands dramatically, reaching nearly $7000$ km, significantly larger than the mere 
$1400$~km obtained in the DM-free case. The latter effect can be regarded as a consequence of the reduction in the threshold mass $M^*$ for electron capture, 
which occurs due to the presence of DM in the white dwarf. The detailed values of all relevant quantities are listed in Table~\ref{tab:tab1}.

We extend the analysis from the $m_{\chi} = 1$~GeV case to $m_{\chi} = 10$~GeV, as shown in the lower panel of Fig.~\ref{fig:M_R_rho} and in the lower part 
of Table~\ref{tab:tab1}.
It can be seen that for $m_{\chi} = 10$~GeV, even a tiny addition of DM leads to significant changes in 
the mass–central-density ($M_{tot}$ -- $\rho_{cOM}$) curve. 

In particular, the central density for the Chandrasekhar mass configuration exceeds the neutron drip density for a DM number fraction as low as $y_{\chi} = 10^{-4}$.  
To account for this, we extrapolate the white dwarf EoS  (Eq.s \eqref{eq:energy_density_2} -- \eqref{eq:Pl}) to large densities.  
Although this introduces some uncertainty, the threshold density for electron capture remains unchanged, and therefore the WD matter EoS remains 
valid for the purpose of locating the electron-capture point. 
For $m_{\chi} = 10$~GeV, the threshold mass $M^*$ at the electron-capture density reduces significantly to $0.718~\mathrm{M_\odot}$ for $y_{\chi} = 5\times 10^{-5}$ 
and $0.370~\mathrm{M_\odot}$ for $y_{\chi} = 1\times 10^{-4}$.  
While the Chandrasekhar mass $M_{Ch}$ decreases with increasing DM fraction, the corresponding radius increases, contrary to the trend seen in the $m_{\chi} = 1$~GeV case. This opposite behavior may be due to the fact that we have extrapolated the white dwarf EoS into a high-density regime where its accuracy is uncertain. However, since our primary interest lies in the threshold mass $M^*$ and corresponding radius at the electron-capture threshold density, 
the EoS for the ordinary white dwarf matter remains valid for our purposes. As in the previous case, the radius at the electron-capture threshold mass increases, and the decrease in threshold mass is more pronounced compared to the $m_{\chi} = 1$~GeV case. While the choice of DM number fractions here is somewhat arbitrary, these results clearly demonstrate the potential influence of DM on white dwarf structure and hint at important implications for electron-capture supernovae, which we will explore in a later subsection.

To gain deeper insight into the impact of asymmetric DM on electron capture processes in DM admixed white-dwarfs, and specifically on the threshold gravitational mass $M^*$ for electron capture, we present in Fig.~\ref{fig:density-profile} the radial mass-density profiles $\rho_j(r)$  of these stars.    
In particular we plot $\rho_j(r)$ for ordinary white dwarf matter (solid lines) and DM (dash-dotted lines) in the case of an 
ordinary ($y_\chi = 0$) white dwarf (solid blue line) and for a DM-admixed white dwarf.   
Both configurations, in each panel of Fig.~\ref{fig:density-profile}, have the same total gravitational mass, equal to the threshold mass for 
electron capture at a given DM number fraction $y_\chi$, i.e. we compare the mass-density profiles for stars having  
$M_{\text{tot}}(y_\chi=0) = M^*(y_\chi = \text{const})$.
The left panel is relative to the case $m_\chi = 1\,\mathrm{GeV}$ and $y_\chi = 3\times 10^{-2}$ which results in $M^* = 1.035 \mathrm{M_\odot}$. 
The right panel is relative to the case $m_\chi = 10\,\mathrm{GeV}$ and $y_\chi = 5\times 10^{-5}$ which results in $M^* = 0.718 \mathrm{M_\odot}$. 

Our results in Fig.~\ref{fig:density-profile} clearly demonstrate that, for the considered values of the DM particle mass $m_\chi$, 
the DM distribution forms a compact core, with a characteristic radius of $R_{DM} \sim 40\,\mathrm{km}$ for $m_\chi = 1\,\mathrm{GeV}$ and 
$R_{DM} \sim 0.4\,\mathrm{km}$ for $m_\chi = 10\, \mathrm{GeV}$. 
The gravitational pull of this dense DM core leads to a significant increase in the central density of ordinary matter, $\rho_{cOM}$, 
and induces a substantial overall compression of the ordinary WD matter fluid (compare the blue and red curves in  Fig.~\ref{fig:density-profile}). 
This compression of the OM fluid is most significant in the region where the DM core exists. Outside the core, the OM density returns to values comparable 
to those in the absence of DM, suggesting that such DM-admixed white dwarfs could still appear consistent with many traditionally observed white dwarfs. For a fixed gravitational mass and a DM number fraction of 
$y_\chi = 3 \times 10^{-2}$, the central density of the OM
component is found to increase substantially in the presence of a DM admixture. 
In the case of $m_\chi = 1\,\mathrm{GeV}$, the OM central density increases from 
$\rho_{c,\mathrm{OM}} \simeq 4.8 \times 10^{7}\,\mathrm{g\,cm^{-3}}$ in the DM-free 
configuration to $\rho_{c,\mathrm{OM}} \simeq 6.8 \times 10^{9}\,\mathrm{g\,cm^{-3}}$, 
corresponding to an enhancement of approximately 2.1 orders of magnitude. 
For $m_\chi = 10\,\mathrm{GeV}$, the OM central density rises from 
$\rho_{c,\mathrm{OM}} \simeq 7.3 \times 10^{6}\,\mathrm{g\,cm^{-3}}$ to the same value of 
$\rho_{c,\mathrm{OM}} \simeq 6.8 \times 10^{9}\,\mathrm{g\,cm^{-3}}$, yielding an enhancement 
of nearly 3.0 orders of magnitude.  The DM component itself attains considerably higher densities. Specifically, 
for $m_\chi = 1\,\mathrm{GeV}$ the central density is 
$\rho_{c,\chi} \simeq 1.6 \times 10^{12}\,\mathrm{g\,cm^{-3}}$, while for 
$m_\chi = 10\,\mathrm{GeV}$ it reaches $\rho_{c,\chi} \simeq 2.1 \times 10^{16}\,\mathrm{g\,cm^{-3}}$. 
These values exceed the central density of the corresponding DM-free white dwarf by roughly 
five to nine orders of magnitude, thereby demonstrating the extreme compactness of the 
DM core relative to the baryonic component.

\subsection{Transition from a DAWD-like Core to a DANS Remnant}

Building on the previously established neutron star model and nuclear matter equation of state (see Subsection~\ref{sect:NS_EOS}) we now examine the 
evolution (collapse) of an electron degenerate DAWD-like core at the electron capture threshold mass $M^*(y_\chi)$ into the DANS remnant via an ECSN 
(hereafter, the DAWD-to-DANS stellar transition process). 

In the scenario we consider in the present work, we assume that during this process both the total baryon number $N_B$ and the total DM particle number $N_D$ are conserved. In other words, we assume that no accretion or ejection of either ordinary matter and DM components occurs during the DAWD-to-DANS stellar transition process. Our scenario thus extend to the case of DM admixed compact stars the scenario proposed by \citet{Bombaci-Datta_2000} to describe the conversion of  a neutron star to a strange star. 

The assumption of baryon number conservation is justified in the context of ECSNe. 
These events are believed to undergo relatively gentle and symmetric collapse, triggered primarily by electron captures on Ne and Mg nuclei, 
and accompanied by limited mass ejection \citep{Nomoto_1984,Nomoto1987,Jones_2013,Zha_2022}.  As a result, the total baryon content of the WD-like stellar core 
is approximately preserved.  
The retention of DM during collapse is supported by the expectation that non-annihilating, weakly interacting DM particles — once gravitationally captured — remain bound to the stellar core due to their coupling to the gravitational potential. Since such particles do not interact significantly with ordinary matter, their spatial distribution and number are largely unaffected by the hydrodynamic and thermal processes that govern the collapse \citep{Goldman_1989, Lavallaz_2010, Kouvaris_2010}. This makes our assumption of constant $N_B$ and $N_D$ across the DAWD-to-DANS transition a reasonable approximation.


\begin{table*}[ht!]
\centering
\resizebox{\textwidth}{!}{%
\begin{tabular}{ccccccccccccc}
    \hline
   \hline
$y_{\chi}$ 
& $M^*_{\mathrm{OM}}$ & $R^*_{\mathrm{OM}}$ 
& $M^*_{\mathrm{DM}}$ & $R^*_{\mathrm{DM}}$ 
& $M^*$ 
& $M^{\mathrm{NS}}_{\mathrm{OM}}$ & $R^{\mathrm{NS}}_{\mathrm{OM}}$ 
& $M^{\mathrm{NS}}_{\mathrm{DM}}$ & $R^{\mathrm{NS}}_{\mathrm{DM}}$ 
& $M^{\mathrm{NS}}_{\mathrm{tot}}$ 
& $E^{conv}$ \\
& [$\mathrm{M_\odot}$] & [km] 
& [$\mathrm{M_\odot}$] & [km] 
& [$\mathrm{M_\odot}$] 
& [$\mathrm{M_\odot}$] & [km] 
& [$\mathrm{M_\odot}$] & [km] 
& [$\mathrm{M_\odot}$] 
& [$\times 10^{53}$ erg] \\
\hline
    \multicolumn{12}{c}{No DM} \\
    \hline
    0     & 1.359 & 1440.7  & & & 1.359 & 1.253&  12.317 & & &1.253& 1.903\\[1ex]
    
    \multicolumn{12}{c}{$m_{\chi} = 1$ GeV} \\
    \hline
    $3  \times 10^{-2}$     & 1.0020 & 3970.193  &0.0333 &40.7306 & 1.0353&0.9385 &  12.1383 & 0.0329 &  5.7617&0.971& 1.144\\
    $6  \times 10^{-2}$     & 0.5582 & 6865.600  &0.0382  &38.6725 &0.5964 &0.5379&  12.1418 & 0.0378 &6.6232 & 0.576&0.357\\
    \\[1ex]
    
    \multicolumn{12}{c}{$m_{\chi} = 10$ GeV} \\
    \hline
     $5 \times 10^{-5}$  & 0.7180  & 7272.626  & $3.854 \times 10^{-4}$ &0.3883 & 0.7184 &0.6905 &   12.612 & $3.853 \times 10^{-4}$ &0.3617 & 0.692& 0.472\\
    $1 \times 10^{-4}$   & 0.3703  & 10399.74 & $3.975 \times 10^{-4}$ & 0.3837&0.3707 &0.3643 &   13.453 &$3.970 \times 10^{-4}$ &0.3666 & 0.365&0.102
    \\
    \hline
\end{tabular}%
}
\caption{ 
Comparison of threshold masses and radii of DM admixed white dwarfs ($M^*$, $R^*$) with the corresponding gravitational masses  and radii of DM admixed neutron stars ($M^{\mathrm{NS}}$, $R^{\mathrm{NS}}$) for various DM number fractions $y_\chi$, for two fixed DM particle masses $m_\chi = 1$ GeV and $m_\chi = 10$ GeV. The values assume constant baryon number across the DAWD-to–DANS transition. 
The columns of the table represent: gravitational mass of ordinary matter at EC threshold configuration ($M^*_{\mathrm{OM}}$), radius of ordinary matter at EC threshold ($R^*_{\mathrm{OM}}$), DM mass at electron-capture threshold ($M^*_{\mathrm{DM}}$), radius of DM distribution at EC threshold ($R^*_{\mathrm{DM}}$), total gravitational mass at the EC threshold ($M^*$), gravitational mass of ordinary matter in the neutron star remnant ($M^{\mathrm{NS}}_{\mathrm{OM}}$), radius of ordinary matter in the neutron star remnant ($R^{\mathrm{NS}}_{\mathrm{OM}}$), DM mass in the neutron star remnant ( $M^{\mathrm{NS}}_{\mathrm{DM}}$), radius of DM in the neutron star remnant ($R^{\mathrm{NS}}_{\mathrm{DM}}$), total gravitational mass of neutron star remnant ($M^{\mathrm{NS}}_{\mathrm{Total}}$), and total  energy released during the DAWD-to-DANS transition ($E_{conv}$).
}
\label{tab:wd_ns_transition}
\end{table*}



In Fig.~\ref{fig:wd_evl_m_rho_R_1} we plot the stellar sequences for DAWD-like cores (continuous lines) and DANS configurations (dashed lines) at fixed and equal 
DM number fraction $y_\chi$ (curves with the same color). The diamond symbol on each of the DAWD curves represents the electron capture threshold mass configuration $M^*(y_\chi)$. The corresponding diamond symbol (connected by the nearly horizontal dash-dotted line) on the DANS sequence with the same $y_\chi$ represents 
the DANS remnant having the same baryon number $N_B$ (and thus the same DM particle number $N_D$) of the initial EC threshold  mass DAWD configuration. 
All the results depicted in Fig.~\ref{fig:wd_evl_m_rho_R_1}  are relative to the case $m_\chi = 1\, \mathrm{GeV}$.    
Similar results, relative to the case $m_\chi = 10\, \mathrm{GeV}$, are drawn in Fig.~\ref{fig:wd_evl_m_rho_R_10}.  
          
Some of the key structural properties for the initial EC threshold mass DAWD configuration and for the final DANS remnant (having the same $N_B$ and $N_D$) are reported in Table~\ref{tab:wd_ns_transition} for different fixed values of the DM number fraction $y_\chi$ and for the two considered values of $m_\chi$. 
In the last column of Table~\ref{tab:wd_ns_transition} we also report the total energy ($E^{conv}$) which is released in the DAWD-to-DANS conversion.  
In the context of our scenario, the stellar conversion energy can be obtained \citep{Bombaci-Datta_2000} as the difference between the total gravitational mass 
$M^* = M^*_{OM} + M^*_{DM}$ of the DAWD (at the EC threshold) and the total gravitational mass $M^{NS}_{Total}$ of the resulting DANS, having the same $N_B$ and $N_D$\, :
\begin{equation}
   E^{conv} =\big( M^* - M^{NS}_{tot}\big)\, c^2\,.
\label{E^conv}    
\end{equation} 
The majority of this energy is expected to be carried away by neutrinos emitted during the electron capture processes in the DAWD-like core that initiate the ECSN, as well as during the subsequent deleptonization phase \citep{Bombaci_1996,Prakash_Phys_Rep_1997} of the proto-DANS.  

In Figure~\ref{fig:dens_profile_WD-NS} we show the mass-density profiles for ordinary matter (solid curves) and for DM (dash-dot curves), 
for the initial EC threshold mass DAWD configuration (left panels) and for the final DANS remnant (right panels). 
The upper panels of Fig.~\ref{fig:dens_profile_WD-NS} correspond to the case $m_\chi = 1\,\mathrm{GeV}$, while the lower panels correspond to  $m_\chi = 10\,\mathrm{GeV}$. 
Thus our Fig.~\ref{fig:wd_evl_m_rho_R_1}, Fig.~\ref{fig:wd_evl_m_rho_R_10}, Table~\ref{tab:wd_ns_transition} and Figure~\ref{fig:dens_profile_WD-NS}
provide complete quantitative information of the initial and final configurations in the DAWD-to-DANS conversion process within our assumptions.  

To highlight some of the central results of our study, we begin by analyzing the baseline scenario of a standard ECSN, characterized by a vanishing DM admixture ($y_\chi = 0$).  
As reported in Table~\ref{tab:wd_ns_transition}, the computed EC threshold mass for the  $^{20}$Ne WD-like core is $1.359\, \mathrm{M_\odot}$, leading to the formation of a Neutron star remnant with a gravitational mass of $1.253\, \mathrm{M_\odot}$, and an associated energy release of $1.903 \times 10^{53}\,\mathrm{erg}$. 
These results are broadly consistent with detailed hydrodynamic simulations of ECSNe incorporating neutrino transport \citep{Zha_2022}. 
In these simulations, which model the collapse of an ONeMg core, a proto-neutron star with baryonic mass of $1.359\,\mathrm{M_\odot}$ and radius of approximately 
$30\,\mathrm{km}$ at $\sim 400\,\mathrm{ms}$ after core bounce is predicted. For a cold, $\beta$-equilibrated neutron star with the same baryonic mass, 
the corresponding gravitational mass is $\sim 1.23\,\mathrm{M_\odot}$ \citep{Zha_2022}. The total energy released — primarily via neutrino emission — is thus on the 
order of $\sim 2 \times 10^{53}\,\mathrm{erg}$, in good agreement with our results. This consistency supports the reliability of our baseline model for standard ECSN evolution in the absence of DM.


As discussed in Subsection~\ref{DM in WD} the presence of DM significantly reduce the EC threshold gravitational mass $M^*(y_\chi)$. 
For example, in the case $m_\chi = 1\,\mathrm{GeV}$ and for a DM number fraction of $y_\chi = 6 \times 10^{-2}$, we find $M^* = 0.596\,\mathrm{M_\odot}$ and   
a resulting DANS remnant with a gravitational mass $M^{NS}_{tot} = 0.576\,\mathrm{M_\odot}$. 
This trend becomes even more pronounced for higher ADM particle masses: for $m_\chi = 10\,\mathrm{GeV}$, we find that even a ADM number fraction 
of $y_\chi = 10^{-4}$ produces a DANS with a gravitational mass of just $0.365\,\mathrm{M_\odot}$. In all cases considered in this study, the radius $R^{NS}_{OM}$ of the ordinary matter distribution in the DANS remnant remains largely unchanged compared to that of a neutron star without DM $R^{NS}_{OM}(y_\chi=0) = 12.317\, \mathrm{km}$ (see Table~\ref{tab:wd_ns_transition} and Figure~\ref{fig:dens_profile_WD-NS} right panels). 

One must also account for the fact that, a key requirement for any neutron star EOS is that it must yield a maximum neutron star mass of at least $2\,\mathrm{M_\odot}$, in accordance with astrophysical observations. Referring to Figures~\ref{fig:wd_evl_m_rho_R_1} and \ref{fig:wd_evl_m_rho_R_10}, one can observe that the neutron star branch for $m_\chi = 1\,\mathrm{GeV}$ becomes inconsistent with this requirement for sufficiently large ($y_\chi \gtrsim 3\times 10^{-2}$) ADM number fractions when using the BL EOS. In contrast, the $10\,\mathrm{GeV}$ case remains fully consistent across all considered values of $y_\chi$. This behavior is also evident in the analysis by \citet{SCORDINO2025371} (see their Fig. 7), using the same BL EOS employed in the present work. 
In particular, for the case $m_\chi = 1\,\mathrm{GeV}$, a DM number fraction of $y_\chi = 6 \times 10^{-2}$ already results in a violation of the $2\,\mathrm{M_\odot}$ constraint, rendering it astrophysically inconsistent. Therefore, the minimum neutron star mass attainable in the $m_\chi = 1\,\mathrm{GeV}$ case  lies between $1.253\,\mathrm{M_\odot}$ (corresponding to $y_\chi = 0$) and $0.971\,\mathrm{M_\odot}$, for $y_\chi = 3 \times 10^{-2}$. On the other hand, for $m_\chi = 10\,\mathrm{GeV}$, 
all $y_\chi$ values used in this study remain consistent with observational constraints, allowing for stable DANS remnants of ECSNe with gravitational masses as low as $0.365\,\mathrm{M_\odot}$.

It should be noted that in all the cases examined in our work, the spatial distribution of DM forms a core both in the initial and in the final configuration of the DAWD-to-DANS transition process (see Figure 5 and Table 2). 
For example in the case $m_\chi = 1\,\mathrm{GeV}$ and $y_\chi = 3 \times 10^{-2}$, we obtain $R^*_{DM} = 40.73\, \mathrm{km}$ and 
$R^{NS}_{DM} = 5.76\, \mathrm{km}$ in other words the DM stellar core shrinks by about a factor of 7 during the DAWD-to-DANS transition. 
Thus in this case the DM core undergoes a moderate collapse  due to its gravitational coupling with the collapse of ordinary matter fluid in the DAWD-DANS process. 
In the case $m_\chi = 10\,\mathrm{GeV}$ the DM core has essentially the same radial distribution (see Fig.~\ref{fig:dens_profile_WD-NS} lower panels) 
and the same radius (see Table~\ref{tab:wd_ns_transition}) both in the initial and final configurations of the DAWD-to-DANS transition.  
Thus, for massive DM particles, the DM distribution remains largely unaffected by the hydrodynamic and thermal processes that govern 
the collapse of the ordinary matter fluid.   
Nevertheless, as we have demonstrated, the presence of DM significantly influences the mass of the DANS remnant and the energetics of the ECSN, 
as we discuss below. 

An important observational consequence of ADM involvement is the substantial reduction in the energy released during the DAWD-to-DANS transition. 
While ECSNe are already associated with low-energy explosions, our results suggest that ADM could further suppress the conversion energy. Therefore, unusually low-energy ECSNe may serve as astrophysical signatures of ADM, offering a potential means to constrain DM properties through multi-messenger observations.


 {
It is also relevant to briefly examine the sub-GeV regime, for which we performed calculations at the representative value $m_\chi=0.1\ \mathrm{GeV}$, in order to assess whether such light DM candidates could leave any distinctive imprint on the WD$\rightarrow$NS transition. To obtain any noticeable impact in this regime, one has to assume unrealistically large DM admixtures, with $y_\chi$ values approaching 0.6 (corresponding to $f_\chi \gtrsim 14\%$). At such high fractions, the WD still has a  DM core at the EC threshold, while the NS remnant invariably hosts a DM halo. The EC--threshold WD masses are (OM+DM (Total)): $1.25582+0.05776\ (1.31358)\,\mathrm{M_\odot}$ for $y_\chi=0.3$, and $1.1602+0.18675\ (1.34695)\,\mathrm{M_\odot}$ for $y_\chi=0.6$. The corresponding NS remnants are $1.1654+0.0572\ (1.2226)\,\mathrm{M_\odot}$ and $1.0837+0.18612\ (1.26982)\,\mathrm{M_\odot}$. The associated conversion energies are $1.63\times10^{53}\ \mathrm{erg}$ and $1.38\times10^{53}\ \mathrm{erg}$, only marginally smaller than the no-DM baseline ($\sim 1.903\times10^{53}\ \mathrm{erg}$). Interestingly, both the WD and the NS remnants display a non-monotonic dependence of their total masses on $y_\chi$: the WD mass first decreases and then increases, while the NS totals follow a similar but weaker trend. Moreover, comparing the spatial distribution of the DM fluid across particle masses shows that in the sub-GeV case the DM extends to large radii, whereas for $m_\chi \sim 1\ \mathrm{GeV}$ the DM core contracts significantly, and for $m_\chi \sim 10\ \mathrm{GeV}$ the core becomes extremely compact. Overall, despite these structural changes, the final NS masses remain close to the no-DM case, and the modest reduction in released energy indicates that sub-GeV ADM might not produce energetically distinctive ECSNe.}

\vspace{1cm}
\noindent  While there is a sustained effort to identify neutron stars across a broad mass spectrum, current observations clearly indicate the existence of massive neutron stars with gravitational masses exceeding $2\,\mathrm{M_\odot}$, which places stringent constraints on the underlying EOS.
 However, an increasing number of observations point toward the existence of neutron stars with significantly lower masses, particularly in X-ray binaries and double neutron star (DNS) systems. Notable examples include 4U~1538--52 ($0.87 \pm 0.07\,\mathrm{M_\odot}$), SMC~X-1 ($1.04 \pm 0.09\,\mathrm{M_\odot}$), and Her~X-1 ($1.07 \pm 0.36\,\mathrm{M_\odot}$) \citep{Rawls_2011, ozel_2012}. In DNS systems, several companions show masses well below the canonical value of $1.4\,\mathrm{M_\odot}$, including PSR~J0453+1559 ($1.174 \pm 0.004\,\mathrm{M_\odot}$), the lowest robustly measured neutron star mass thus far, PSR~J1756--2251 ($1.230 \pm 0.007\,\mathrm{M_\odot}$) \citep{Ferdman_2014}, and PSR~J0737--3039B ($1.249 \pm 0.0007 \,\mathrm{M_\odot}$) \citep{Kramer_2006}.  Additionally, emerging systems like PSR~J1946+2052 show companions with minimum mass estimates around $1.2\,\mathrm{M_\odot}$ \citep{Stovall_2018}, reinforcing the growing evidence for a population of neutron stars with significantly lower masses than those expected from standard core-collapse supernovae. These low-mass neutron stars challenge the conventional paradigm of iron-core collapse supernovae and instead point toward alternative formation channels, such as electron-capture supernovae (ECSNe)  resulting from the collapse of  (O–Ne–Mg) white dwarfs. Theoretical models predict that such processes lead to gravitational masses in the range of $1.15$--$1.25\,\mathrm{M_\odot}$ \citep{Nomoto1982, Nomoto1987, Nomoto_1984, Lattimer_2012}. Our study supports this formation pathway, demonstrating that the observed low-mass neutron stars, compactness constraints, and white dwarf progenitor properties are consistent with ECSN origins. These findings reinforce the astrophysical relevance of WD $\rightarrow$ NS evolution and open new avenues for constraining the low-mass end of the neutron star mass distribution and the microphysics of stellar collapse.

\section{Summary and Conclusion} \label{sec:summary}


Electron-capture supernovae (ECSNe) have been proposed as a robust formation channel for low-mass neutron stars, supported by both theoretical modeling and recent observations \citep{Hiramatsu2021}. Interestingly, both the ECSN explosion mechanism and the compact remnants it leaves behind, such as 
low-mass neutron stars \citep{Jones_2016} or ONeFe white dwarfs \citep{Jones_2019}, may serve as valuable probes of dark-sector physics. 
Recent studies suggest that DM or dark photons could influence the evolution, collapse dynamics, and observational signatures of these systems, offering a novel window into the properties of dark-sector particles under extreme conditions \citep{PhysRevLett.134.151002}. Motivated by these developments, in this work we investigated the impact of fermionic asymmetric DM (ADM) on ECSNe and the formation of low-mass neutron stars, employing a general relativistic two-fluid formalism. 
In this model, ordinary matter (OM) and DM (DM) are treated as separately conserved fluids, interacting solely through gravity. We focused specifically on progenitor cores modeled as neon-rich white dwarfs (Ne WDs), incorporating the electron-capture process self-consistently using a Gibbs free energy threshold criterion.

For white dwarf matter, we adopted a zero-temperature equation of state (EOS) with Coulomb lattice corrections suitable for the crystallized ionic phase, while neglecting minor corrections from electron exchange and polarization effects. The progenitor cores were assumed to be composed primarily of $^{20}$Ne, motivated by their critical role in ECSN events. The neutron star remnants were described using the Brueckner–Hartree–Fock (BHF) EOS, developed from chiral effective field theory ($\chi$EFT) interactions, ensuring consistency with empirical nuclear matter properties and astrophysical constraints from gravitational wave observations.

To explore the evolution from ADM-admixed white dwarfs (DAWDs) to ADM-admixed neutron stars (DANSs), we assumed conservation of baryon number ($N_B$) and DM particle number ($N_D$) during the stellar collapse. This assumption is justified due to the minimal mass ejection typical of ECSNe and the negligible non-gravitational interactions between ADM and ordinary matter. Solving the two-fluid Tolman–Oppenheimer–Volkoff (TOV) equations under these constraints allowed us to predict the properties of neutron star remnants formed through ADM-assisted collapse. However, it should be noted that our analysis is limited to static snapshots before and after collapse, without detailed modeling of the dynamical evolution.

Our results demonstrate that even a modest ADM fraction  (for high enough DM particle mass) significantly increases the central density of white dwarf progenitors and substantially reduces the threshold gravitational mass ($M^*$) required for electron capture to trigger ECSNe. Consequently, ADM presence enables ECSNe from progenitor cores with lower mass, leading to stable neutron stars with gravitational masses potentially well below the minimum neutron star mass observed so far,  thus offering a plausible astrophysical pathway for the formation of ultra-low-mass neutron stars.
For example, in the case of $m_\chi = 10\,\mathrm{GeV}$ and $y_\chi = 10^{-4}$, we find that the resulting DM-admixed neutron star (DANS) can be as light as $\sim 0.36\,\mathrm{M_\odot}$. In contrast, for $m_\chi = 1\,\mathrm{GeV}$, the lowest attainable mass is $\sim 0.97\,\mathrm{M_\odot}$, limited by the requirement that the EOS supports a $2\,\mathrm{M_\odot}$ neutron star, which excludes higher ADM fractions. 
Furthermore, our results indicate that the conversion energy during the DAWD–to–DANS transition decreases notably with higher ADM particle masses and fractions. 
This reduction in conversion energy points to the possibility that low-luminosity ECSNe could serve as indirect signatures of ADM participation in the collapse process.  {However, in the sub-GeV regime ($m_\chi \sim 0.1,\mathrm{GeV}$), the impact on progenitor structure and energetics is minimal, yielding neutron star masses comparable to the no-DM case and suggesting that only heavier ADM can leave distinctive astrophysical imprints.}


We emphasize that while our framework presents a consistent and physically motivated picture, it is based on certain simplifying assumptions that could influence the quantitative outcomes. In particular, we have restricted our analysis to ECSNe originating from white dwarf cores composed of $^{20}\mathrm{Ne}$, assuming idealized conditions. In more realistic scenarios, the progenitor WD-like core is expected to consist of a stratified $^{16}$O, $^{20}$Ne, and $^{24}$Mg composition, with varying mass fractions shaped by prior shell burning and convective mixing. These compositional gradients, along with inhomogeneities in the electron fraction, can influence the onset of electron captures and alter the critical density and temperature at which dynamical collapse is triggered, thereby affecting both the threshold mass for collapse and the energetics of the ensuing supernova. On the neutron star side, uncertainties in the high-density EOS, including the possible presence of  exotic degrees of freedom, may further modify the predicted minimum neutron star mass and total released energy in the ECSN. Additionally, the maximum allowed ADM fraction could be affected by more sophisticated treatments of DM transport or baryon, DM interactions. Despite these limitations, the qualitative trends observed in our results remain stable under plausible physical variations. 

In conclusion, this study represents a first step toward modeling ADM-triggered ECSNe within a relativistic two-fluid framework that incorporates electron capture. It provides a physically grounded perspective on the potential role of DM in stellar collapse and compact object formation, and motivates future work that integrates more detailed stellar evolution modeling, nuclear microphysics, and multi-messenger observational inputs.

\bibliography{biblio}{}
\bibliographystyle{elsarticle-harv}



\end{document}